\documentclass[12pt]{iopart}
\usepackage{iopams}
\usepackage{iopams,cite}  
\usepackage{eurosym}
\usepackage{lipsum}
\usepackage{graphicx}
\usepackage{color}
\usepackage{amssymb}
\usepackage[normalem]{ulem}
\usepackage{lipsum}
\usepackage{tikz} 
\usepackage{multicol}
\usepackage{float}
\begin{document}
\title{Vortex patterns in rotating dipolar Bose-Einstein condensate mixtures with squared optical lattices}
\author{Ramavarmaraja Kishor Kumar$^{1}$, Lauro Tomio$^{2,3}$, and Arnaldo Gammal$^{1}$}
\address{$^1$Instituto de F\'{i}sica, Universidade de S\~{a}o Paulo, 05508-090 S\~{a}o Paulo, Brazil \\
$^2$Instituto de F\'{i}sica Te\'{o}rica, Universidade Estadual Paulista, 01140-700 S\~{a}o Paulo, Brazil \\
$^3$Instituto Tecnol\'ogico de Aeron\'autica, DCTA,12.228-900 S\~ao Jos\'e dos Campos, Brazil}

\begin{abstract}
Vortex lattice patterns with transitions from regular to other variety vortex shapes are predicted 
in rotating binary mixtures of dipolar Bose-Einstein condensates loaded in squared optical lattice. 
We focus our investigation in the experimentally accessible dipolar isotopes of dysprosium ($^{162,164}$Dy), 
erbium ($^{168}$Er), chromium ($^{52}$Cr),  and rubidium ($^{87}$Rb), by considering
the binary mixtures ($^{164}$Dy-$^{162}$Dy, $^{168}$Er-$^{164}$Dy,  $^{164}$Dy-$^{52}$Cr and $^{164}$Dy-$^{87}$Rb),
which are confined in strong pancake-shaped trap and loaded in squared two-dimensional optical lattices, 
where we vary the polarization angle of dipoles, the inter-species contact interactions and the rotation frequency.  
The ratio between inter- to intra-species contact interaction is used for altering the miscibility properties; with  
the polarization of the dipolar species used for tuning to repulsive or attractive the dipole-dipole interactions.
For enough higher rotation, of particular interest is the regime when the inter- to intra-species scattering length is larger 
than one, in which a richer variety of vortex-lattice patterns are predicted, including vortex sheets and two-dimensional 
rotating droplet formations. The patterns can be controlled by changing the optical lattice parameters, as shown for the
symmetric $^{164}$Dy-$^{162}$Dy dipolar mixture. For mixtures with stronger differences in the dipole moments, as 
$^{164}$Dy-$^{52}$Cr and $^{164}$Dy-$^{87}$Rb, only half quantum vortices and circular ones have been observed, 
which will depend on the dipole orientations.
\end{abstract}
\maketitle
\section{Introduction}
Bose-Einstein condensate (BEC) systems loaded in optical lattice potentials are quite relevant in the investigation of fundamental problems 
in condensed matter physics~\cite{Bloch2008}, such that different accessible experimental possibilities are of interest to be studied.
For instance, by considering a gas of ultracold atoms with repulsive interactions 
held in a three-dimensional optical lattice, a quantum phase transition from superfluid to a Mott insulator
was observed as the potential depth of the lattice is increased~\cite{Greiner2002}. 
For a considerable large number of atoms localized in single wells of an optical lattice (OL), it can be more safely applied
the mean-field theory. When there are few number of atoms in each lattice, the phase coherence between 
the condensed atoms in the different lattice sites can be lost with the system undergoing a Mott-insulating phase 
transition~\cite{Fisher1989}. 
By exploring the phase coherence of coupled quantum gases in two-dimensional (2D) optical lattices, the lifetime of the 
condensate and the dependence of the interference pattern on the lattice configuration have been investigated in 
Ref.~\cite{Greiner2001}.
Also, BECs loaded in OL can be used as testing grounds for strongly correlated condensed matter systems~\cite{Ottl2005}. 
The phase transition between superfluid-Mott insulator was observed experimentally in a rubidium $^{87}$Rb cold-atom 
gas~\cite{Bakr2010}. 

Mixtures of atomic BECs with different species can provide a broader range of  possibilities to study quantum phenomena. 
Since 1997, there have been experimental investigations with two-component 
mixture of hyperfine states of $^{87}$Rb~\cite{Myatt1997}. Further, different heteronuclear 
atomic species have been combined to produce BEC mixtures. 
The mass imbalance between the heteronuclear species and their particular intra- and interspecies interaction 
and the selective tuning of the intra- and interspecies interaction have allowed investigating several interesting 
phenomena. Recent experimental progress with $^{168}$Er and $^{164}$Dy condensates, as well as with 
their mixtures are challenging and stimulating present investigations on the properties of dipolar 
BECs~\cite{2016Ferrier,2016Schmitt,2016Chomaz,2018Ferrier,2018Chomaz,2018Ilzhofer,2018Trautmann}.  
Experiments reported  on stability and collapse of single component dipolar BECs loaded in OL  can be
found in Refs.~\cite{2011Muller}. Also, single component dipolar gases in two-dimensional OL have been 
studied theoretically by using the Bose-Hubbard model and mean-field approximation, from where   
new quantum phases are predicted, as checkerboard or supersolid phases~\cite{2009Lahaye,2012Baranov}. 

From experiments reported in Ref.~\cite{2006Tung}, vortices pinning in rotating BECs have been observed,
in which an orientation locking between the vortices and the optical lattices was verified.
By considering the effects of 2D triangular and squared optical lattices, it was also shown that sufficiently high 
squared optical lattice will induce a structural crossover in the vortex lattice.
In this Ref.~\cite{2006Tung}, it is also explained how vortex nucleations depend on the depth of the optical lattice
potential and on the rotation frequency. Later, in Ref.~\cite{2010Williams}, it was demonstrated the existence of a 
linear dependence of the number of vortices with respect to the rotation frequency in the case of deep OL.
It is common to observe square vortex lattices in binary mixtures of rotating BECs in harmonic 
trap~\cite{2003-Kasamatsu,Kumar2017}. But, a single BEC loaded in a rotating OL can show a rich variety of vortex structural 
transitions, including square vortex structures~\cite{Reijnders2004,Pu2005,Kasamatsu2006}, when playing with other 
parameters of the condensed atomic gas. 
An increasing depth of the OL potential helps to reduce the critical rotation frequency necessary to appear a 
single vortex. This potential is considered deep when it is greater than the chemical potential of the system. 

Previous studies dealing with vortices in dipolar BECs have explained the role of dipole-dipole interactions (DDI) in the formation of
vortices~\cite{2017Martin}. In particular, the theoretical analysis dealt with the calculation of the critical rotation frequency and vortex 
structures under the action of the DDI~\cite{Yi2006,Malet2011,Kishor-vort}. 
Two-component dipolar BECs in pancake-type traps  
have shown to feature several types of vortex lattices, which can be observed by controlling parameters such as the rotation frequency, 
the ratio between inter- to intra-species scattering lengths, as well as the magnetic moment orientations of the dipoles.
In this regard, by considering opposite polarized two-component BEC with strong DDI in three dimensions, a
 dynamical study on stability and pattern formation was recently reported in Ref.~\cite{2018Xi}.
In these kind of studies, the addition of OL potentials is also expected to show some relevant effects in the vortex-lattice structures,  
which are worthwhile to be investigated.  
In this direction, we notice some recent theoretical simulations in Refs.~\cite{2016Zhang}, by considering OL effects in BEC with 
the dipolar atomic elements $^{52}$Cr, $^{164}$Dy and $^{168}$Er. 
These works, besides being limited to OL  in just one of the directions of the pancake-type harmonic trap, are already 
indicating that rich structures can be obtained by extending the OL in a symmetric way to the two directions of the 2D trap. 
It is also of interest to explore the region of parameter space, by varying the inter- to intra-species scattering ratios, which can 
be performed via Feshbach techniques~\cite{1998inouye}, as well as dipolar parameters by also studying other experimentally 
accessible dipolar condensed mixtures.  For a general approach to compute vortex-lattice in rotating condensed systems with 
strong repulsive two-body interaction and lattice potentials, we can also point out the Ref.~\cite{2009Zeng}.

Another property that plays a relevant role in dipolar mixtures is the miscibility of two-component BEC 
systems~\cite{Chui-1998,2012Liu,2013Pattinson}.
This property was recently studied by considering the atomic isotopes $^{162,164}$Dy and $^{168}$Er~\cite{2017jpco}.  
With respect to miscibility of dipolar binary mixtures, the rotational properties differ and show distinct vortex lattice 
structures~\cite{Kumar2017}. In view of more recent experiments on tuning the magnetic DDI using a dysprosium 
condensate, we understand as quite appealing to investigate the properties of BECs with different coupled dipolar 
species. In particular, it is appropriate to include the dysprosium as one of the coupled species in investigations 
with dipolar mixtures, in view of its large dipole moment~\cite{2011Lu}.

In the present work, our main interest is to provide a more extended investigation on vortex-lattice structures,
focusing in binary dipolar mixtures of coldatoms which are accessible for experimental realization,
exploring the possibilities to tune the DDI under rotating squared optical lattices.
This study can have particular interest to indicate region of parameters which can be tuned in experimental investigation,
as well as in the analysis of density patterns of binary dipolar mixtures, 
in which some new unusual soliton formations and/or quantum lattice phases can emerge~\cite{Tang2018}.
In addition to the previous binary mixtures that have been studied in Refs.~\cite{2017jpco,Kumar2017} 
($^{164}$Dy-$^{162}$Dy and $^{168}$Er-$^{164}$Dy), here we also discuss two cases of coupled systems
that are less-symmetric  in their DDI as well as in their respective masses, which are  
$^{164}$Dy-$^{52}$Cr and  $^{164}$Dy-$^{87}$Rb. 
For the mixture with $^{87}$Rb, which is an atomic element with quite weak magnetic dipole,
we found not necessary to explore in details, once is verified that the corresponding relevant results follow 
from cases such as with $^{52}$Cr, where one of the dipole moment is not strong. 
By analyzing mixtures that are partly and completely immiscible, we show a variety of vortex structures that are 
controlled by larger interspecies contact interactions.  
Besides the fact that an increasing depth of the OL potential helps to reduce the critical rotation frequency 
to produce vortices, in our study we assume this potential is not too deep, as limited by the validity of the 
mean-field model.
Supported by our numerical simulations, we provide phase diagrams indicative of the expected vortex-lattice patterns,
in a plane defined by the rotation frequency $\Omega$ and the ratio between inter- and intra-species scattering lengths 
$\delta$ for two different orientation angles of the dipoles, such that the DDI can be repulsive or attractive.

In the next Sec. 2, we present the formalism that we consider for the treatment of coupled dipolar condensed mixtures, 
confined by pancake-shaped harmonic trap with rotating optical lattice. In Sec. 3, we present our results for the different kind of coupled
mixtures we are considering, together with details on the numerical approach and corresponding parameters. 
Finally, in Sec.4 we summarize the main findings with our conclusions. 

\section{Binary dipolar condensate under rotation loaded in squared optical lattice}
\label{sec2} 
In our present study of coupled two-species dipolar condensates, our main objective is to investigate a rotating condensate under 
the effect of a squared optical lattice, which is acting together with a strong  pancake-shaped 
harmonic trap in the system. This matter is quite demanding to be theoretically studied in view of recent experimental 
interests with condensed quantum gases with DDI and rotating magnetic fields~\cite{Tang2018}. 
By tuning the alignment angle of the two-dipole species, the DDI interaction can be changed from repulsive to attractive.
 For the three-dimensional (3D) DDI kernels, we assume the relative positions of the two dipole moments makes an
angle $\theta$ relative to a defined $z-$direction. With that, by considering the condensate in a pancake geometry 
defined in the $(x,y)-$plane, with the relative positions between the dipoles being near $90^\circ$, we have mainly a 
repulsive DDI for a static magnetic field fixed in the $z-$direction. 

For the DDI, we follow the scheme as described in  Ref.~\cite{goral2002,2002Pfau}, with the polarization of both aligned dipoles 
being under rotation due to a time-dependent rotating external magnetic field.  
So, we consider the orientations of both dipoles making an angle $\varphi$ with respect to the $z-$axis, such that the 
DDI is repulsive when $\varphi=0$, becoming attractive for $\varphi>54.7^\circ$.
As detailed in Ref.~\cite{2002Pfau}, the tunability of the magnetic dipolar interaction is performed by using 
time-dependent magnetic fields with dipoles rapidly rotating around the $z-$axis. The magnetic field is given by the 
combination of a static part along the $z-$direction and a fast rotating part in the $(x,y)-$plane, having a frequency 
which is chosen such that the atoms are not significantly moving during each period. Under these conditions, 
once performed a time averaging of the DDI in a period, the corresponding 3D averaged interaction 
for the coupled two dipolar species, with the magnetic dipole moments $\mu_1$ and $\mu_2$ given 
in terms of the Bohr magneton $\mu_B$, can be written as
\begin{equation}
\left<V^{(d)}_{3D}({{\bf r}_1-{\bf r}_2})\right>=
{\mu_{0}\mu_{1}\mu _{2}}
\frac{1-3\cos ^{2}\theta}{\left|{{\bf r}_1-{\bf r}_2}\right|^{3}}
\left(\frac{{3\cos ^{2}\varphi}-1}{2}\right), 
\label{DDI}
\end{equation} 
where $\mu _{0}$ is the free-space permeability. The factor within parenthesis in (\ref{DDI}) is the result of the
time-averaging procedure on the dipole orientation around the $z-$axis.  
The angle $\varphi$ provides the effective strength and sign of the interaction, with the magnetic dipoles being 
completely polarized along the $z-$direction for $\varphi=0$ (when $V_{ij}^{(d)}>0$), with the sign being inverted 
for $\varphi\ge 54.7^\circ$ when the interaction becomes attractive.  
It is worthwhile to notice in Eq.~(\ref{DDI}) that the DDI effect can be cancelled out at two specific orientations:
when the polarization is such that $\cos^2\varphi=1/3$, or when the relative position of the two dipoles in relation to the 
$z-$axis is  such that $\cos^2\theta=1/3$. However, in our following approach we assume $\theta\sim 90^\circ$, as we consider 
the binary system in strongly pancake-shaped format.

One of our present aims is to study the effect of a squared optical lattice potential in dipolar mixtures under
rotation. This OL potential is added together with the external 3D harmonic trap, which is confining both atomic species $j=1,2$ 
in a strong pancake-shaped symmetry, given by the trap aspect ratio $\lambda=50$, such that 
\begin{equation}
V({\bf r})=\frac{1}{2}m_j\omega_j^2(x^{2}+y^{2}+\lambda^2 z^2) + V_{ol} \left[\sin^2(kx)+\sin^2(ky)\right],
\label{3Dtrap}
\end{equation}  
where $m_j$ and $\omega_j$ are, respectively, the mass and trap frequency of the species $j$. 
The strength of the OL potential is given by $V_{ol}$, with $\pi/k$ defining the
lattice periodicity which is assumed to be identical in the $x$ and $y$ directions.

The strong pancake-shaped symmetry ($\lambda=50$) assumed along this study, is considered due to
stability requirements. 
With such strong asymmetry, together with the symmetry of the optical lattice potential and the 
DDI, the corresponding 3D coupled equation is reduced to 2D format, by the
usual factorization of the 3D wave function $\Psi _{j}(\mathbf{r},t)$ into the ground state of the transverse 
harmonic oscillator trap and a 2D wave function~\cite{Luca,Luca2,CPC2,2012-Wilson,Kumar2017}, with
\begin{equation}
\Psi _{j}(\mathbf{r},t)=\left(\frac{\lambda}{\pi l^2}\right)^{1/4}e^{-{\lambda z^{2}}/{(2l^2)}}\; \Phi _{j}(x,y,t). \label{ansatz1}
\end{equation}
Next, we perform the 2D reduction by introducing the above ansatz in the original 3D Gross-Pitaevskii (GP) 
formalism with DDI, integrating over the variable $z$.
The final coupled equations are cast in a dimensionless format by measuring the energy  in units of 
$\hbar \omega_{1}$, with the length in units of $l\equiv \sqrt{\hbar/(m_{1}\omega _{1})}$.
For this purpose, the space and time variables are given in units of $l$ and $1/\omega_1$, respectively, with 
${\bf r}\to l {\bf r}$ and $t\to \tau/\omega_1$.  Within this procedure, the corresponding dimensionless
wave-function components are given by $\psi_{j}(x,y,\tau)\equiv l \Phi _{j}(x,y,t)$.
The two-body contact interactions related to the scattering lengths $a_{ij}$, as well as the dipole-dipole interaction 
parameters for the species $i,j = 1,2$ are defined as
\begin{eqnarray}
g_{ij}&\equiv& \sqrt{2\pi \lambda }\frac{(m_{1}+m_2)}{m_2}\frac{a_{ij}N_{j}}{l},\;\;
\mathrm{d}_{ij}=\frac{N_{i}}{4\pi }\frac{\mu_{0}\mu_{i}\mu _{j}}{\hbar \omega _{1}\,l^{3}},
\nonumber\\
a_{ii}^{(d)}&\equiv& \frac{1}{12\pi }\frac{m_{i}}{m_{1}}\frac{\mu_{0}\mu _{i}^2}{\hbar \omega _{1}l^{2}},
\;\; a_{12}^{(d)}=a_{21}^{(d)}=\frac{1}{12\pi }\frac{\mu_{0}\mu_{1}\mu _{2}}{\hbar \omega _{1}l^{2}},
\label{par}
\end{eqnarray}
where $N_{j=1,2}$ is the number of atoms in the species $j$. 
With the assumption that both components are confined by pancake-shaped harmonic traps 
with the same aspect ratio $\lambda$ and such that the relation between the corresponding trap 
frequencies is given by  $m_2\omega_2^2\simeq m_1\omega_1^2$,  
in terms of the above notation for the units and parameters, the corresponding 
coupled 2D GPEs (with $i,j=1,2$ and $j\ne i$) can be written as 
{\small \begin{eqnarray}
\mathrm{i}\frac{\partial \psi_{i}  }{\partial \tau }
&=&\,{\bigg [}-\frac{m_{1}}{2m_{i}}{\left(\frac{\partial^2}{\partial x^2}+\frac{\partial^2}{\partial y^2}\right)}
+V(x,y)- \Omega L_{z}+
\sum_{j=1,2}g_{ij}|\psi_{j} |^{2}
\nonumber\\
&+&\sum_{j=1,2}  \mathrm{d}_{ij} \int_{-\infty}^{\infty}d x^\prime d y^\prime V^{(d)}_{2D}(x-x^\prime,y-y^\prime)
|\psi_{j}^\prime  |^{2}
 {\bigg ]}
\psi_{i}  
\label{2d-2c}, 
\end{eqnarray}}
where $V^{(d)}(x,y)$ is the 2D expression for the DDI, which is to be derived from the 
corresponding 3D counterpart given in Eq.~(\ref{DDI}), and 
$\psi_{i}\equiv \psi_{i}(x,y,\tau)$ and $\psi_{i}^\prime\equiv \psi_{i}(x^\prime,y^\prime,\tau)$
are the components of the total 2D wave function, normalized to one, 
$\int_{-\infty}^{\infty}dxdy|\psi _{i}|^{2}=1.$
$L_z$ is the angular momentum operator with $\Omega$ the corresponding rotation parameter (in units of  $\omega_1$), 
which is common for two components.
In this 2D reduction, the external potential provided by the harmonic trap, together with the 
squared optical lattice, in dimensionless units is given by 
\begin{equation}
V(x,y)=
\frac{1}{2}(x^{2}+y^{2}) + V_0 \left[\sin^2(kx)+\sin^2(ky)\right],\label{trap}
\end{equation} 
where $V_0\equiv V_{ol}/\hbar\omega_1$ gives the strength of the OL in dimensionless units, 
with $k$ being the laser wave vector (also dimensionless and being the same in both $x$ and $y$ directions). 

The 2D DDI is presented in Eq. (\ref{2d-2c}) can be expressed by means of the convolution theorem, which
is given by the inverse 2D Fourier-transform for the product of the DDI and densities, as 
\begin{equation}
\int dx^\prime dy^\prime V^{(d)}(x-x^{\prime },y-y^{\prime })|\psi _{j}^{\prime}|^{2}
=\mathcal{F}_{2D}^{-1}\left[ \widetilde{{V}}^{(d)}(k_x,k_y){\widetilde{n}_{j}}(k_x,k_y)\right],
\end{equation}
where $ {\widetilde{n}_{j}}(k_x,k_y)$ is the 2D Fourier transform of the density,  
\begin{eqnarray}
 {\widetilde{n}_{j}}(k_x,k_y)
 &=& \int dx dy\; e^{{\rm i}(k_xx+k_yy)} |\psi _{j}  |^{2}  
, \label{2Ddens} 
\end{eqnarray}
with $\widetilde{V}^{(d)}(k_x,k_y)$ being  the Fourier transform of the DDI, which is expressed by the 
combination of two terms, corresponding to parallel and perpendicular polarizations of the dipoles in relation to the 
$z-$direction~\cite{Ticknor2011}:
\begin{eqnarray}
\widetilde{V}^{(d)}(k_x,k_y)&\equiv& \cos^2(\varphi) \widetilde{V}_\perp^{(d)}(k_x, k_y) 
+ \sin^2(\varphi) \widetilde{V}_\parallel^{(d)}(k_x, k_y).
\label{V2D}
\end{eqnarray}
The DDI is perpendicular when the two dipoles are polarized in the $z-$direction ($\varphi=0$), 
\begin{eqnarray}
\widetilde{V}_\perp^{(d)}( k_x, k_y) &=& 2 - 3 \sqrt{\pi} \left( \frac{k_{\rho}}{\sqrt{2\lambda}}\right)
\exp \left( \frac{k_{\rho }^{2}}{2\lambda}\right) {\rm erfc}\left( \frac{k_{\rho }}{\sqrt{2\lambda}}\right)\label{DDI-perp},
\end{eqnarray}
and longitudinal when the dipoles are polarized in the $(x,y)$ plane ($\varphi=90^\circ$),  
\begin{eqnarray}
\widetilde{V}_\parallel^{(d)}(k_x, k_y) &=& -1 + 3 \sqrt{\pi}\left( \frac{k_{\rho}\cos^2\theta_k}{\sqrt{2\lambda}}\right)
\exp \left( \frac{k_{\rho }^{2}}{2\lambda}\right) {\rm erfc}\left( \frac{k_{\rho }}{\sqrt{2\lambda}}\right)\label{DDI-par}.
\end{eqnarray} 
where $k_\rho\equiv\sqrt{k_x^2+k_y^2}$, 
with $\theta_k$ being  
an arbitrary direction in the momentum plane defined by $(k_x,k_y)=(k_\rho\cos\theta_k,k_\rho\sin\theta_k)$.
As all $\theta_k$ are equally possible, by averaging $\cos^2\theta_k$ we obtain a factor 1/2.
So, the total DDI is averaged to zero for $\varphi=54.7^\circ$, being repulsive for $\varphi<54.7^\circ$ 
and attractive for  $\varphi>54.7^\circ$. As the system trapped in an optical lattice potential can become unstable 
when the dipolar BEC is tuned to larger angles, the attractive part of the DDI is restricted to  $\varphi\le 60^{\circ}$. 

\section{Vortex pattern results: dipolar mixtures in optical lattices  } 
\label{sec3}
Our main results are reported in this section, by considering different kind of 
coupled dipolar systems.  An isotope of the dysprosium was chosen as a common element in the 
mixtures in view of its large magnetic moment, as well as due to the fact that this element was 
considered in recent experimental BEC realizations in ultra-cold laboratories. We select a few
sample results indicating the rich variety of vortex-lattice patterns, which are obtained with dipolar 
mixtures under rotation with squared optical lattices. 
For the more illustrative cases of symmetric- and asymmetric-dipolar mixtures, we present diagrams 
considering the rotation parameter as a function of the ratio between inter- to intra-species contact
interactions, for repulsive and attractive DDI. These diagrams, which are summarizing the kind of vortex-lattice 
patterns that we have obtained, are followed by characteristic examples.

Before reporting the specific dipolar systems we have obtained, next we provide some details 
on the corresponding physical parameters and numerical approach.

\subsection{\bf Parameter space and numerical approach}

Of particular relevance to the present study on dipolar binary mixtures is to consider atomic species accessible to 
experimental realizations in ultracold atom laboratories, which have significant magnetic dipole moments. By 
considering the dipole moments of each species, the corresponding dipolar parameter of the DDI will be fixed 
for each specific mixture. In this regard, for the binary mixtures we are using, the corresonding magnetic dipole 
moments given in terms of the Bohr magneton $\mu_B$, are the following:
$\mu=10\mu_B$ for $^{162,164}$Dy, $\mu=7\mu_B$ for $^{168}$Er, $\mu=6\mu_B$ for $^{52}$Cr and
$\mu=1\mu_B$ for $^{87}$Rb.
In particular, considering the corresponding dipole moments, the strengths of the DDI are given as 
$a_{ij}^{(d)}=131\,a_{0}$ ($i,j=1,2$), for the $^{164}$Dy-$^{162}$Dy mixture; 
$a_{11}^{(d)}=66\,a_{0} $, $a_{22}^{(d)}=131\,a_{0}$ and $a_{12}^{(d)}=a_{21}^{(d)}=94\,a_{0} $, 
for the $^{168}$Er-$^{164}$Dy mixture; 
and $a_{11}^{(d)}=131\,a_{0} $, $a_{22}^{(d)}=16\,a_{0}$ and $a_{12}^{(d)}=a_{21}^{(d)}=25\,a_{0} $, 
for the $^{164}$Dy-$^{52}$Cr mixture.
Once fixed the DDI, in order to explore various families of vortex patterns we vary 
the rotation frequency $\Omega$, the polarization angle $\varphi$, as well as the ratio $\delta$ between 
inter- and intra-species contact interaction. 

Apart from dipolar parameters, other intrinsic properties of binary systems are related to the two-body contact 
interactions, which in principle can be varied by using Feshbach techniques~\cite{1998inouye}.
In order to keep our study to stable systems, all the two-body scattering lengths are assumed to be positive, 
with the intra-species ones  being identical and given by $a_{11}=a_{22}=40a_0$ ($a_0$ is the Bohr radius), 
with the inter-species one, $a_{12}$, obtained from the ratio $\delta=a_{12}/a_{11}$. 
Relevant for the vortex-patterns that we are studying, we allow $\delta$ to be changed in a region of interest 
in the parameter space, from smaller to larger values, together with the parameter $\Omega$ related to the 
rotation of the condensate and parameters of the squared optical lattice.

The contact and DDI parameters, which appear in the Eq.~(\ref{2d-2c}) as dimensionless ($g_{ij}$ and $d_{ij}$, 
respectively) are given in units of the Bohr radius $a_0$ by (\ref{par}).  By varying these parameters we  
contemplate several conditions of interest in view of miscibility properties of the binary mixtures (see a previous 
study in Ref.~\cite{2017jpco} without optical lattice). 
For the usual harmonic trap, in all the following simulations and analysis of the results we consider 
a strong pancake-shaped trap in the $(x,y)-$plane, with an aspect ratio $\lambda =50$, fixing the number of atoms for 
both species as $N_{1}=N_{2}=10^{4}$. These choices are appropriate for experimental realistic settings due to 
stability requirements.  
The assumed angular frequencies of the harmonic axial traps are such that 
$m_2\omega_2^2\simeq m_1\omega_1^2$, implying that $\omega_j = 2\pi\times 60 {\rm s}^{-1}$ for the $^{168}$Er and 
$\omega_j = 2\pi\times 61 {\rm s}^{-1}$ for the $^{162,164}$Dy.

The external trap is modified by the addition of a squared optical lattice, defined in the same
$(x,y)-$plane. By choosing the spacing of the OL given by $d_{lat}= \lambda_L/2 \approx534$nm, as in the
experiments reported in Ref.~\cite{Muller2011}, the corresponding wavelength of the laser is  
$\lambda_L=1064$nm. So, with the time and space units such that $1/\omega_1=$2.65 ms and $l=1\mu$m,  
the corresponding dimensionless parameters for the OL spacing is 
$ \pi/k = 0.534$; or $k=1.87 \pi$.  
In the results that we are presenting the OL depth was  fixed to $V_0\,=\,15$. However, by investigating other
strengths, going up to $V_0\,=\,25$, we have observed that the vortex structural properties remain 
essentially unchanged.

In order to solve Eq.~(\ref{2d-2c}), we employ  the split-step Crank-Nicolson method, as in Refs.~\cite{CPC2,Gammal2006,CPC1}, 
combined with a standard method for evaluating DDI integrals in the momentum space~\cite{CPC2,goral2002}. In order to search 
for stable solutions, the numerical simulations were carried out in imaginary time on a grid with maximum of $640$ points in 
$x$ and $y$ directions, with spatial steps $\Delta x=\Delta y=0.05$ and time step $\Delta t=0.0005$. 
Both component wave functions are renormalized to one at each time step.
In order to obtain stationary vortex states, we solve the Eq.~(\ref{2d-2c}) with different initial conditions. 
In view of previous tests on the initial suitable conditions, for that we use a combination of angular harmonics, 
followed by verifying the convergence of the solutions for the inputs, as discussed in Refs.~\cite{Kumar2017,Rokhsar,Bao,Jeng}.

\subsection{\bf Symmetric-dipolar mixture $^{164}$Dy-$^{162}$Dy in squared optical lattice}

Our analysis on dipolar mixtures starts by considering the symmetric case with coupled dipolar mixture of
two dysprosium isotopes, $^{164}$Dy and $^{162}$Dy, such that the strengths of the DDI are the same,
given by $a_{ij}^{(d)}=131\,a_{0} $ ($i,j = 1,2$). The main characteristics of this system, verified in previous
studies, is that it is completely miscible as verified in a study without OL, having the same value close to one for the 
miscibility parameter defined in Ref.~\cite{2017jpco}, implying in an almost complete overlap between the densities 
of the two components.  
In view of the miscibility of the two components, the parameter $\delta$, which gives the ratio between inter- and 
intra-species scattering length, plays the main role for the observation of quite different vortex patterns, for a given 
rotation $\Omega$.

\begin{figure}[H]
\vspace{-0.3cm}
\begin{center}
\includegraphics[width=0.47\textwidth]{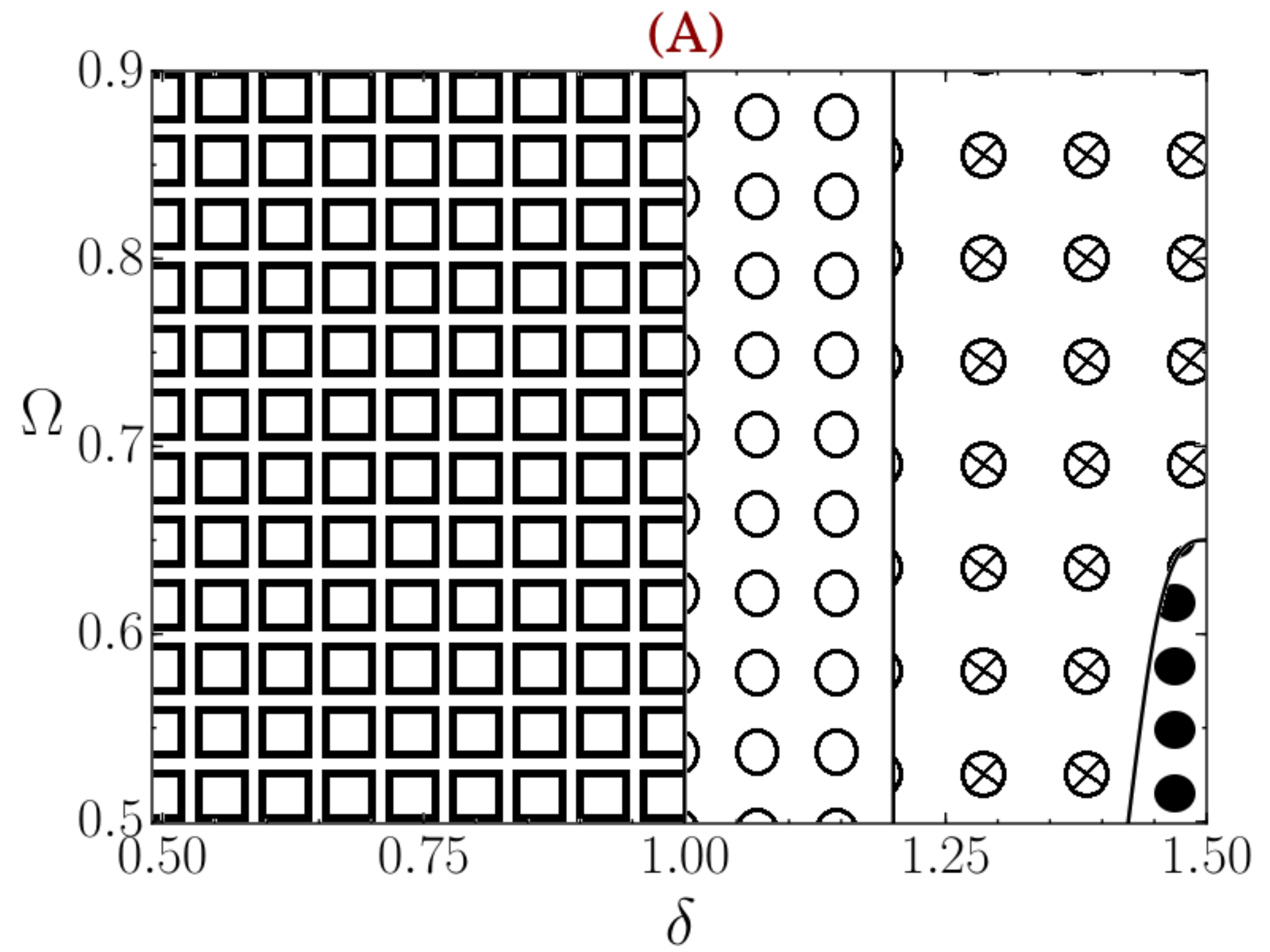}
\includegraphics[width=0.47\textwidth]{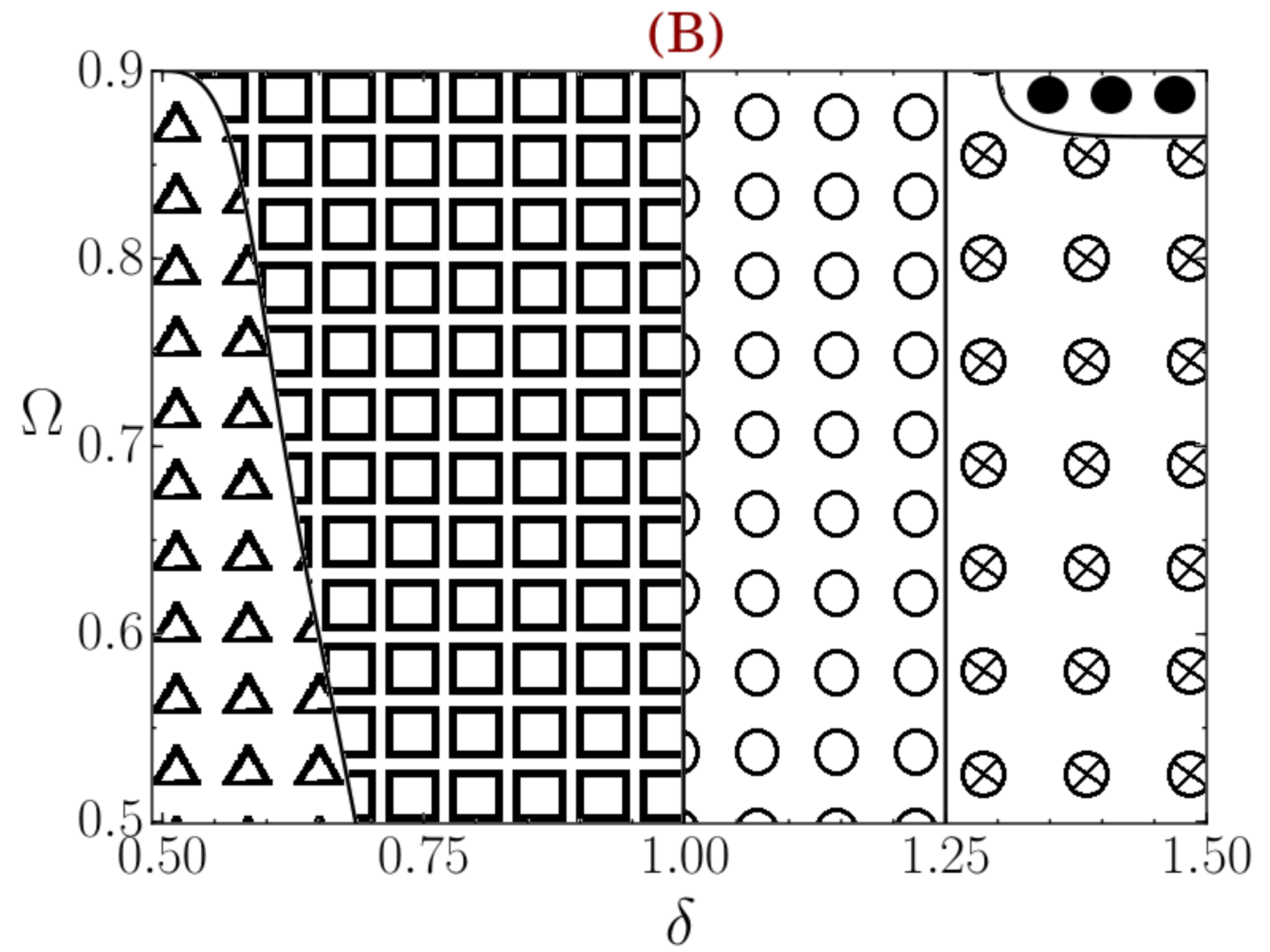}
\end{center}
\vspace{-0.5cm}
\caption{
Diagrammatical representations of vortex-lattice patterns obtained for the $^{164}$Dy-$^{162}$Dy dipolar mixture,

with the rotation frequency $\Omega$ as function of the contact interaction ratio $\delta\equiv a_{12}/a_{11}$

 in a plane defined by $\Omega$ (in units of $\omega_1$) and contact interaction ratio $\delta\equiv a_{12}/a_{11}$. 
 Both dipoles are polarized: along the $z-$axis ($\varphi=0$) in (A); and at an angle $\varphi=60^\circ$ in (B). 
 The symbols filling specific intervals indicate the approximate observed vortex-lattice patterns in these regions: 
 squares for squared lattices; triangles for triangular lattices; circles for double-core or striped vortices; 
 crossed circles for domain walls or vortex sheets; and solid circles for 2D rotating droplets.}    
\label{fig01}
\end{figure}

 In Fig.~\ref{fig01}, by considering our results with different values of $\Omega$ (from
0.5 to 0.9) and $\delta$ (from 0.5 to 1.5), we present two diagrams which are indicating the main characteristic of vortex 
patterns that are obtained for the $^{164}$Dy-$^{162}$Dy dipolar mixture, confined by a strong pancake-type harmonic 
trap with a squared OL (having strength $V_0=15$).
In this figure, we consider two different orientations for the dipoles of the coupled mixture, in relation to the $z-$axis.
In the diagram (A) we have the dipoles oriented parallel to $z$,  such that $\varphi=0^\circ$ for the 
corresponding angle. With this orientation the DDI is repulsive. For the results shown in the diagram (B), the dipole 
polarizations are tuned to an angle $\varphi=60^\circ$ in relation to the $z-$axis, supporting attractive DDI.
In both the panels, for a given region ($\Omega$, $\delta$), the kind of observed vortex patterns are being 
identified by the symbols filling the respective region:
(i) with squares, for squared-vortex lattices; (ii) with empty circles, for double-core or striped vortex patterns;
(iii) with triangles, for triangular-vortex lattices; (iv) crossed circles, for domain-wall or vortex-sheet patterns;
and (v) solid circles, for lattice patterns with quantum rotating droplet formations.
\footnote{For ``rotating droplets" we meant droplet-like density oscillations, 
similar as the ones discussed in Refs.~\cite{2005Kasamatsu,2010revmodphys}. 
In our present study we are not considering quantum fluctuations, such that these droplets cannot be associated 
to the self-bound droplets reported in~\cite{2016Ferrier,2016Chomaz} for single-component experiments.}

As observed for this symmetric mixture, when the dipoles are aligned along the $z-$axis,
for $ \delta\le 1$, the OL is changing the lattice patterns from triangular ones, which have
been verified when no optical lattice is active, to squared lattice patterns.
This kind of patterns has been observed even for $\delta=0.1$, with the 
effect of OL being more evident in the regions where inter-species contact interaction is 
considerably smaller than intra-species interaction. 
As shown in the phase diagram (A), for the case that $\varphi=0$ (when the dipoles are aligned with $z$),
for $1<\delta\lesssim 1.2$ the patterns change to double-core and striped vortices; and domain walls for 
$\delta > 1.2$. These pattern results are almost independent of the value of the frequency parameter 
$\Omega$, which was varied for $0.9>\Omega>0.5$. However, among the patterns with domain walls observed for $\delta > 1.4$,
we have also verified lattice patterns with rotating droplets, in case that $\Omega<0.65$.

By tuning  the dipoles to an angle $\varphi=60^\circ$ in relation to the $z-$axis, we are providing an attractive dipolar 
interaction. In the binary mixture, repulsive two-body interactions determine the miscibility properties. So, the attractive 
dipolar interactions are going to make the system more miscible, reducing the effect of the OL, in particular for smaller 
values of $\Omega$.  Consequently, part of the squared vortex structures observed when $\delta<0.7$ in the phase 
diagram (A) (when $\varphi=0$), will change to  triangular  ones, as shown in the phase diagram (B) of Fig.~\ref{fig01}. 
We can also observe that this attractive dipolar interaction is playing a role in the droplet regime when $\delta>1$, as 
the attractive interactions increase the critical rotation frequency for single vortex.
So, it becomes explicit the reduction in the number of vortices due to the attractive dipolar interaction. 
As the rotation velocity $\Omega<0.85$ is not sufficient to produce the droplet regime, 
the rotating droplets are observed for $\delta>1.3$ and  $\Omega>0.85$. 

Typical examples of pattern results corresponding to diagram (A) of Fig.~\ref{fig01}, for $\Omega=0.5$ and 0.9, are displayed 
in the next Figs.~\ref{fig02} to \ref{fig04}, in which the dipoles are aligned along the $z-$axis ($\varphi=0$). 
In Fig.~\ref{fig05} we have results for attractive DDI ($\varphi=60^\circ$) with rotation frequency $\Omega=0.9$. 

\subsection*{\bf $^{164}$Dy-$^{162}$Dy mixture, with repulsive dipole-dipole interaction}

\begin{figure}[h]
\begin{center}
\includegraphics[width=15cm]{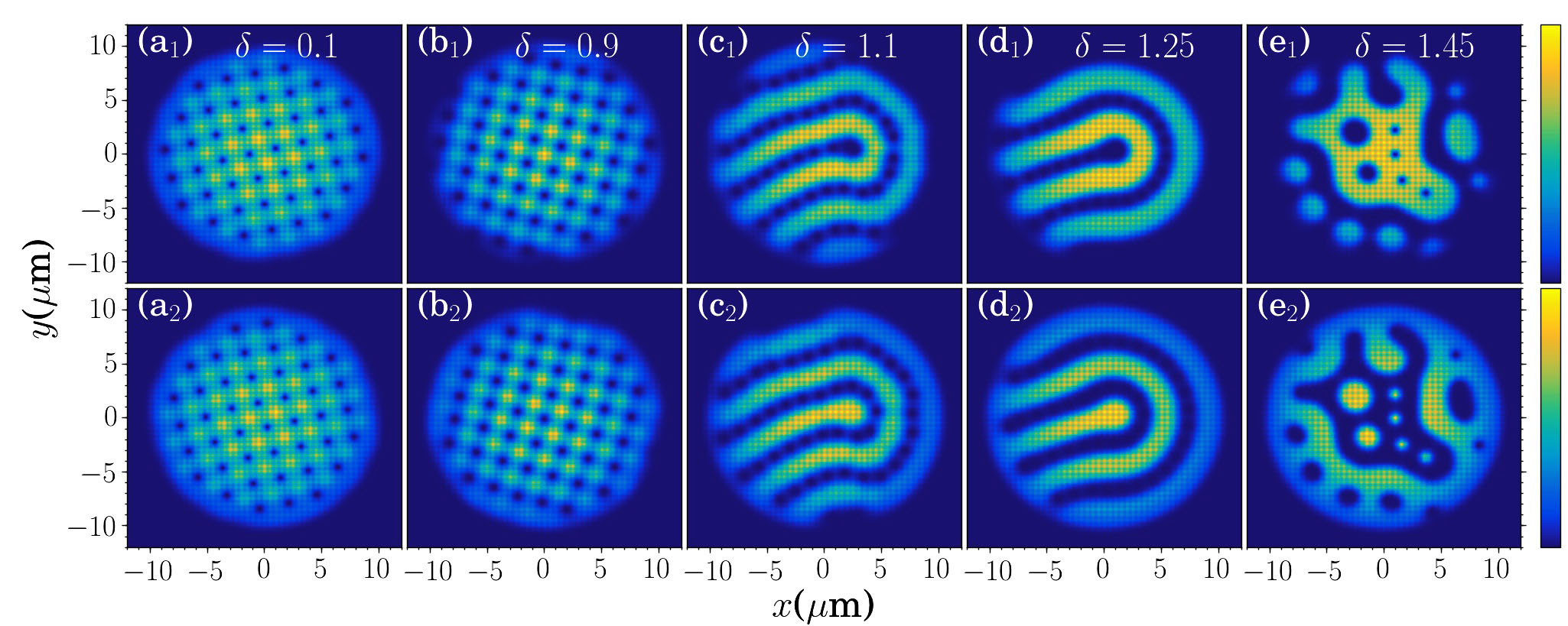}
\end{center}
\vspace{-.5cm}
\caption{
The 2D densities $|\psi_{j}|^{2}$ for the  symmetric-dipolar mixture $^{164}$Dy-$^{162}$Dy 
(when $a^{(d)}_{ij}=131 a_0$, $i,j=1,2$) are shown in the $(x,y)$ plane for different stable lattice-vortex patterns, 
with $0.1<\delta<1.45$ (as indicated inside the panels), considering $\Omega =0.5$ (units of $\omega_1$).
Both dipole orientations are parallel to the $z-$axis ($\varphi=0$, repulsive DDI), with the OL parameters 
being $V_0=$15 ($\hbar\omega_1$ units) and $\pi/k=0.534$ ($\mu$m units).  
The other parameters are $N_j=10^{4}$, $\lambda =50$, and $a_{jj}=40a_{0}$. Starting from zero (darker), 
the maximum density levels (clearer) are 0.009 for ($a_{j}$) and 0.012 for the other panels, in $(\mu$m$)^{-2}$ units.
} \label{fig02}
\end{figure}

For the repulsive case, we start in  Fig.~\ref{fig02} to show the corresponding densities of the two components of the coupled
dipolar system, by considering the ratio between two-body inter- and intra-species $\delta$ scattering lengths changing from 0.1 to 1.45,
which are given in five pairs of panels for the coupled elements of the mixture, in the presence of a squared OL (with $V_0=15$),
for a given rotation frequency $\Omega=0.5$. The effect of the squared OL in this case that we have repulsive DDI can be verified by 
comparing the results with a previous results performed without OL, for $\Omega<0.6$ and $\delta<0.6$. We noticed that vortex-lattice 
patterns with triangular formats observed in Ref.~\cite{Kumar2017} are modified to squared formats, when adding  a squared OL with 
grid size given by $k=1.87\pi$, as exemplified by the panels (a$_{j=1,2}$). 
Further, irrespective to the values of $\Omega$, for $\delta \leq 1$, we have observed only square-shaped vortex structures, as
shown in the panels (a$_{j}$) and (b$_{j}$).
However, when $\delta$ becomes larger than one, with the inter-species contact interactions becoming dominant, the miscibility of the two
components start to play a more relevant role, resulting in lattice-vortex structures more favorable to immiscible bosonic systems.  
These results are represented by the panels (c$_{j}$),  (d$_{j}$) and (e$_{j}$), for $\delta=$1.1, 1.25 and 1.45, respectively. 
The closely located vortices of each species start to join together forming patterns with domain walls separating the species, resulting in
vortex sheets or with serpentine shapes. This behavior in pattern transitions can be seen from (b$_j$) to (c$_j$), and from  (c$_j$) to (d$_j$).

Transition to less regular patterns is observed by enhancing the immiscibility of the mixture, increasing even more the relation between 
inter- to intra-species scattering length $\delta$, as indicated by the panels (e$_j$). The two dipolar species start 
to become almost separated, but sharing the same region of space as they are symmetric in their dipolar properties with
masses slightly different. Therefore, patterns with different shapes can be observed as we increase the value of $\delta$,
where 2D rotating droplet formations can emerge.  This behavior is characterized by the panels (e$_{j}$), in which 
is noticed that the more central region is mainly occupied by the more massive species.

\subsection*{\bf Vortex-lattice changes by varying the optical-lattice spacing.}

\begin{figure}[H]
\vspace{-0.cm}\begin{center}
\includegraphics[width=15cm]{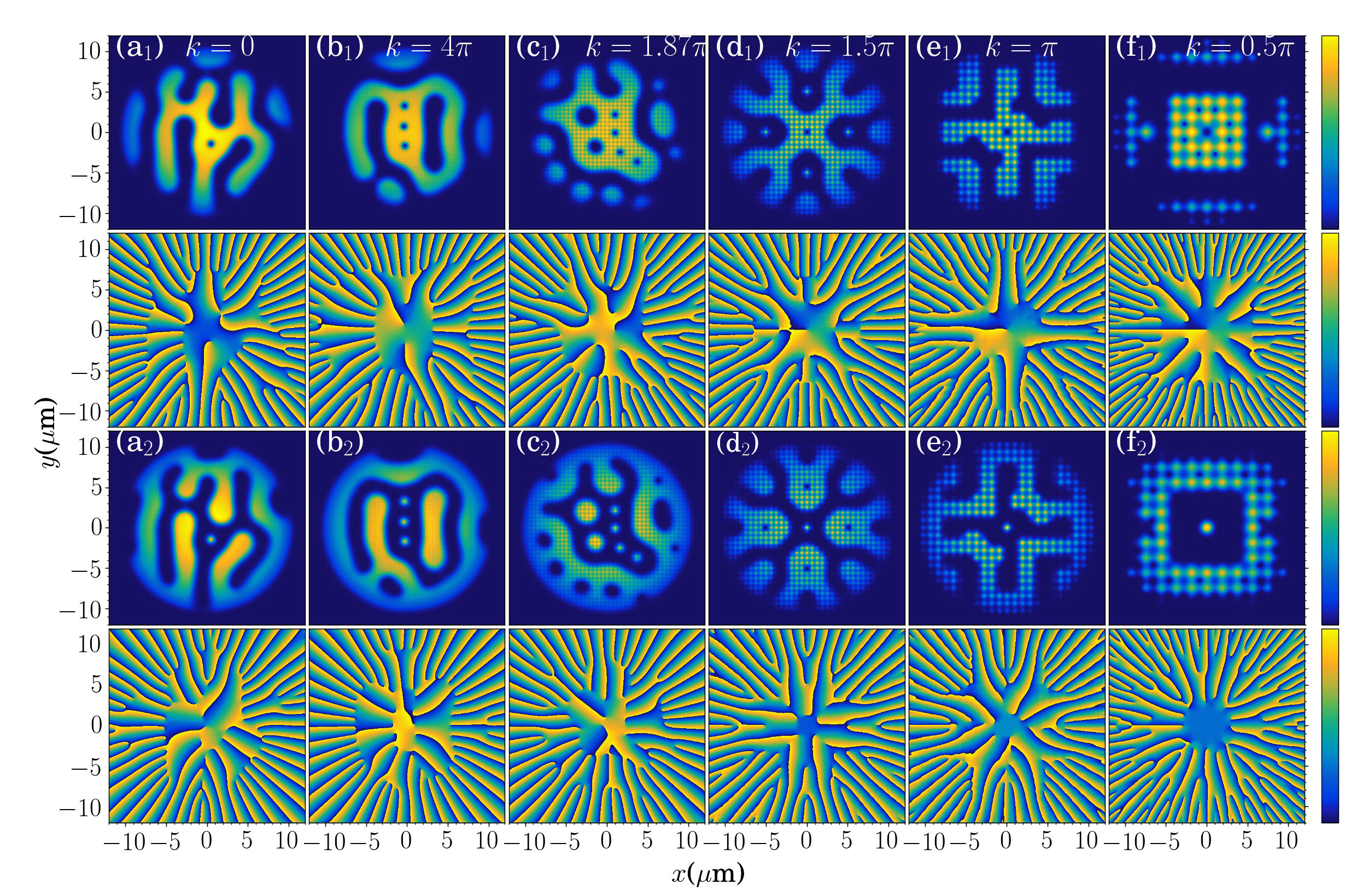}
\end{center}
\caption{
The effect of the OL in the vortex-lattice patterns shown in Fig.~\ref{fig02} is verified by an array of panels
where the densities are given in the 1st and 3rd rows ($j=1,2$, respectively), followed by the corresponding 
phases, in the 2nd and 4th rows. 
For that, we fix $\delta=1.45$ and vary the OL parameter $k$ (OL is switched off for $k=0$), keeping all the other 
parameters as in Fig.~\ref{fig02}.
As shown, the OL potential has stronger impact in the patterns in the interval $0.5\pi \lesssim k \lesssim 2\pi$. 
The mapping levels are from zero (darker) to 0.012 (clearer) for the densities and from $-\pi$ to $\pi$ for the phases.} 
\label{fig03}
\end{figure}

The optical lattice spacing used along this work, $ \pi/k = 0.534$ (corresponding to $k=1.87 \pi$) is motivated by our main 
purpose to investigate an experimentally accessible region for the coupled dipolar systems, as in Ref.~\cite{Muller2011}. 
However, in order to verify the effect of the spacing grid given by the optical lattice parameter $k$ in the pattern transitions, 
we include Fig.~\ref{fig03} by considering a fixed large value of $\delta=1.45$ [the same as in the panels (e$_j$) of 
Fig.~\ref{fig02}], with $k$ varying from zero to $4\pi$; the first implying there is no OL potential, as shown by Eq.~(\ref{trap}); 
with the second having a grid-spacing so small that the  net effect is reduced to a constant in the trap. 
The panels presented in Fig.~\ref{fig03} are arranged in an array with six columns and four rows, by contemplating the 
densities together with their corresponding phases. The first and third rows of panels are presenting the densities 
$|\phi_{j}|^2$ for the first ($j=1$) and second ($j=2$) components of the mixture, respectively; with the second and fourth rows 
presenting the corresponding phases (for the panels identified in the first and third rows). The values of $k$ ($=0, 4\pi$, 
$1.87\pi$, $1.5\pi$, $\pi$ and $0.5\pi$) are fixed at each column, being identified in the first row, such that in the first column
we have the case without OL potential. Next, in the other panels we consider the squared OL grid starting from smaller size 
($\pi/k=0.25$) to larger size ($\pi/k=2$).
In this regime with $\delta=1.45$, when switching off the OL with $k=0$, the patterns are dominated by
the repulsive dipolar interactions and rotation frequency ($\Omega=0.5$), with  the results, for densities and phases, being 
shown in the first column of Fig.~\ref{fig03}. 
By considering very large values of $k$, such as $k=4\pi$ (implying in smaller grid sizes) the results have some similarities
with the ones obtained for the case without OL ($k=0$), 
indicating that the OL effect is averaged out. However, by increasing the grid of the OL potential, as shown in the panels 
with columns identified by (d$_j$) to (f$_j$), we notice that it is more evident the role of the OL when considering 
intermediate values of $k$. With the panels (d$_j$) for $k=1.5\pi$ to (f$_j$) for $k=0.5\pi$, it is observed that
the effect of OL is dominating, resulting in more geometric patterns, whereas near $k=1.87\pi$ 
a transition regime is verified, with formation of 2D rotating droplets in the mixture.

Without OL and symmetric binary mixtures, rotating droplets have already been observed in Ref.~\cite{2003-Kasamatsu},
with instabilities and pattern formations been recently studied in Ref.~\cite{2018Xi}, by considering oppositely 
polarized dipoles in a two-component Bose-Einstein condensate. 
Here, by applying shallow squared optical lattice for the rotation given by $\Omega = 0.5$  2D droplet shaped densities are 
formed in the particular regime with $\delta > 1.4$, as indicated in the phase diagrams of Fig.~\ref{fig01} and  exemplified 
in the panels (d$_j$) of Fig.~\ref{fig02}, as well as in the column identified by (c$_j$) of Fig.~\ref{fig03}.
Due to the immiscibility, rotating droplet like density peaks are formed near to the surface of the BEC in the first component;
being located near the middle of the condensate for the second component. 

For the row of panels with the densities given in the first and third rows of the Fig.~\ref{fig03}, we show the
corresponding phases in the second and fourth rows of the same figure. These plots with the phases are featuring the 
corresponding vorticity, such that we can see the vorticity corresponding to the droplets in the column with (c$_j$).
In the structure of the droplets that are being shown for the phases, we observe about vorticity two in the small-size 
droplets, and about four in the large-size droplets. These structures are similar to the double-core structures, where 
droplets in any one of the components are formed by multiple vortices with the same circulation, 
with vortices in any of those components being surrounded by those multiple vortices.

\begin{figure}[H]
\begin{center}
\includegraphics[width=0.8\textwidth]{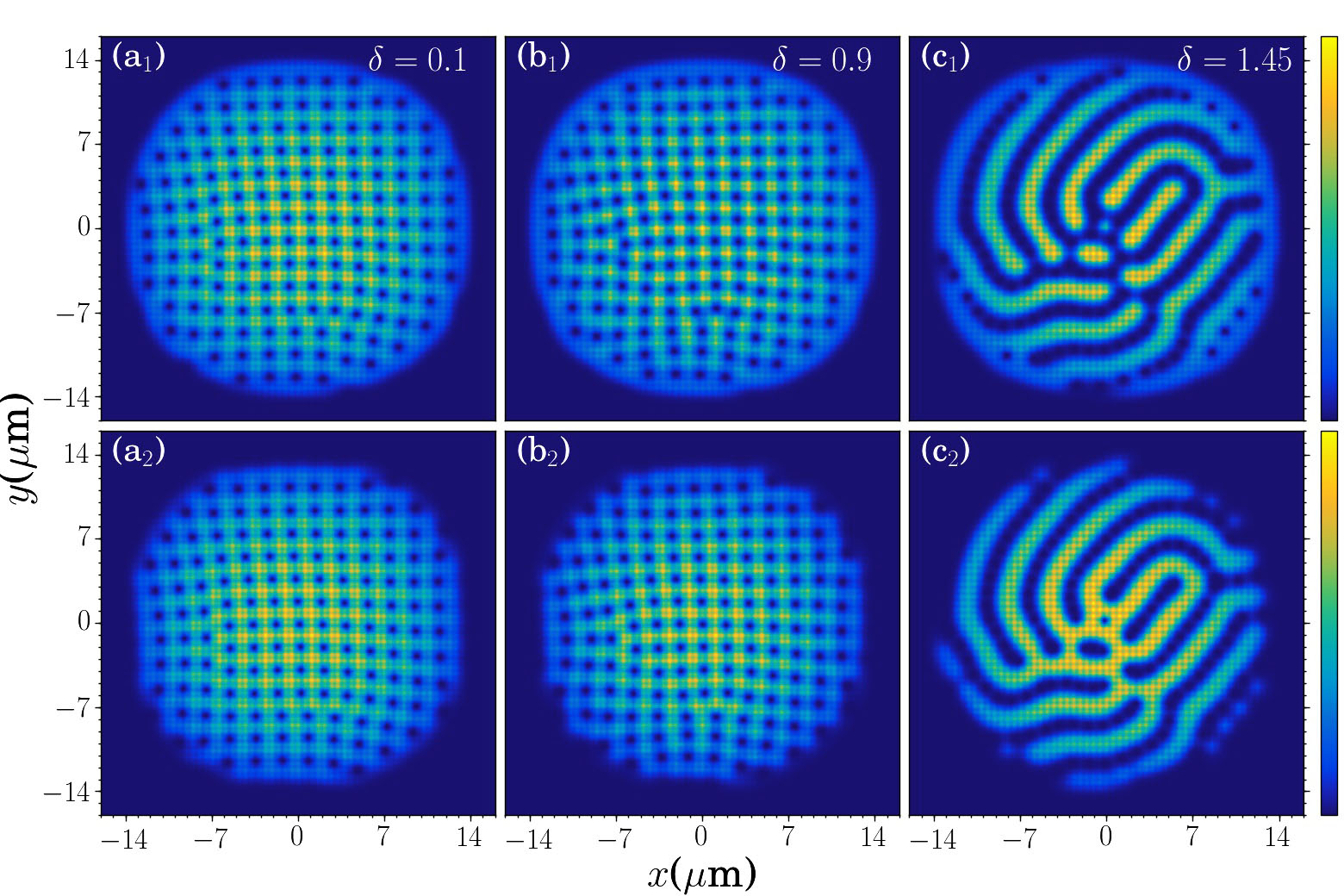}
\end{center}
\caption{Effect of increasing the rotation frequency to $\Omega=0.9$ in the densities 
shown in Fig.~\ref{fig02}, where the DDI is repulsive ($\varphi=0$).
For the components $j=$1 (upper) and 2 (lower), the panels are for $\delta=$ 0.1, 0.9 and 1.45 (indicated inside
the upper panels).
The patterns have squared formats in (a$_{j}$) and (b$_{j}$), with striped and domain-wall structures in (c$_{j}$).  
Other parameters and units are as in Fig.~\ref{fig02}. 
Starting from zero (darker), the maximum density levels (clearer) are 0.0054 for the panels ($a_{j}$) and  
($b_{j}$), and  0.0068 for the panels ($c_{j}$). }
\label{fig04}
\end{figure}

For the same $^{164}$Dy-$^{162}$Dy mixture, in Fig.~\ref{fig04} we are verifying the vortex-pattern structure when 
increasing the rotation parameter, from 0.5 that was used in Figs.~\ref{fig02} and \ref{fig03} to $\Omega=$0.9.
By comparing both the cases given by Figs.~\ref{fig02} and \ref{fig04} in the regime with $\delta > 1.45$,
we noticed that rotating droplets are no more supported and mostly we observe the formations of domain walls in the binary mixture. 
As the rotation becomes stronger, the binary mixture expands in the plane, being distributed in a larger radius.
In this regime with $\delta>1$, the  droplet density peaks change to domain walls due to the large number of vortices, 
which become connected with each other. 
However, as we decrease $\delta$ below one, we return to vortex-lattice structures having square formats. 

\subsection*{\bf $^{164}$Dy-$^{162}$Dy mixture, with attractive dipole-dipole interaction}

\begin{figure}[H]
\begin{center}
\includegraphics[width=0.8\textwidth]{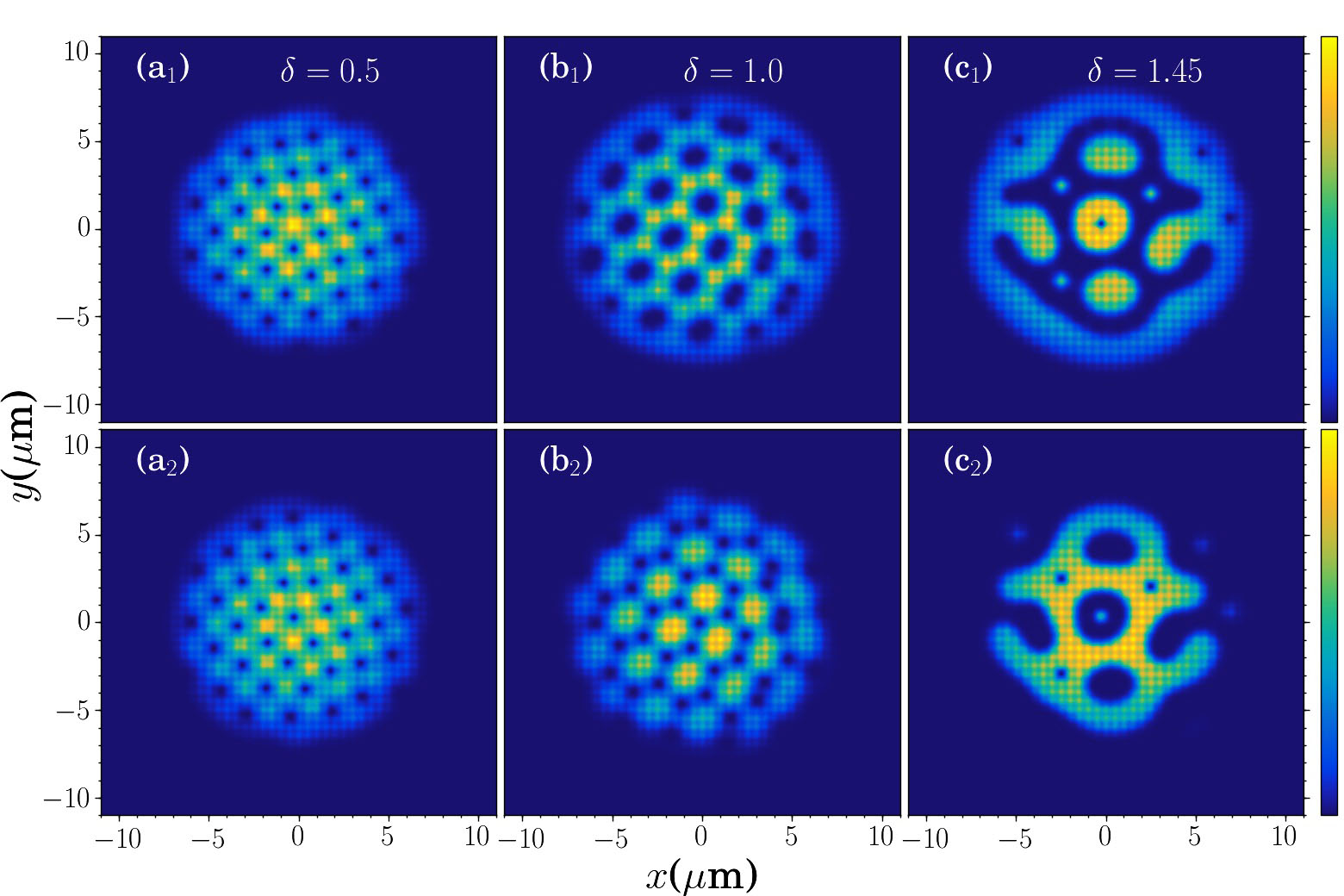}
\end{center}
\caption{
By keeping $\Omega=0.9$, as in Fig.~\ref{fig04}, we show the effect of changing the dipole
orientations to attractive DDI with $\varphi=60^\circ$.
The panels are for $\delta=$0.5, 1.0, and 1.45 (as indicated inside the upper panels), with other parameters as in 
Fig.~\ref{fig04}. By comparing with Fig.~\ref{fig04}, we noticed the vortex-lattice patterns changing from squared 
formats to other geometric formats in (a$_{j}$) and (b$_{j}$) (going to hexagon-like formats) and from striped 
to domain-walls with rotating droplets in (c$_{j}$).} 
Starting from zero (darker), the maximum density levels (clearer) are 0.021 in all the panels.
\label{fig05}
\end{figure}

The orientation of the dipoles is expected to be relevant, considering that by changing the both aligned dipoles from
$\varphi=0^\circ$ to $\varphi=60^\circ$ we are also changing the strength of the DDI from repulsive to attractive one.
Therefore, with the objective to verify how much the vortex-lattice patterns could be affected by changing the
orientation of the dipoles in this symmetric case, we present in Fig.~\ref{fig05} results corresponding to the ones we
have verified in Figs.~\ref{fig02} to \ref{fig04}. Now, we consider the polarization of the dipoles
making an angle $\varphi = 60^\circ$ in relation to the orientation of the $z-$axis, such that the DDI becomes attractive.
In order to compare with previous results, in Fig.~\ref{fig05} we consider the same rotation frequency $\Omega=0.9$ as 
 in Fig~\ref{fig04}. 
Again, we consider three values of the inter- to intra-species scattering lengths, given by $\delta$ (smaller, equal and 
larger than one). The main changes that we observe by comparing the Figs.~\ref{fig04} and \ref{fig05} are that the attraction 
in the DDI reduces the radius from $\sim$ 14 to $\sim$ 6 and, due to the increasing miscibility between the components, 
the patterns observed for $\delta>1$ somewhat similar to the ones verified in Figs.~\ref{fig02} and
\ref{fig03}, with domain walls and with some rotating droplet formations (more restricted in this case due to more attraction 
between the components).

\subsection{\bf Asymmetric-dipolar mixture $^{168}$Er-$^{164}$Dy in squared optical lattice}

\begin{figure}[H] 
\begin{center}
\includegraphics[width=0.4\textwidth]{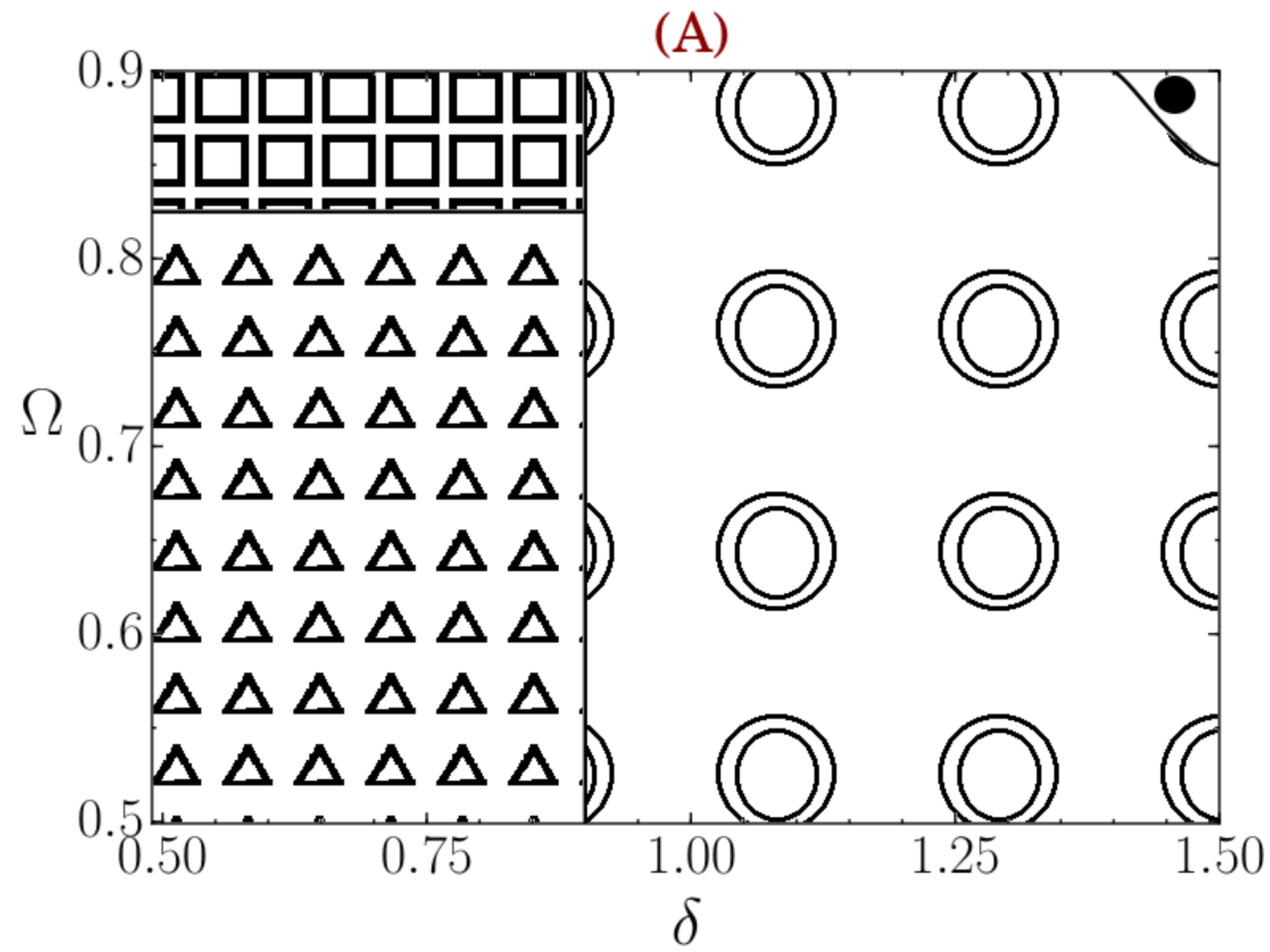}
\includegraphics[width=0.4\textwidth]{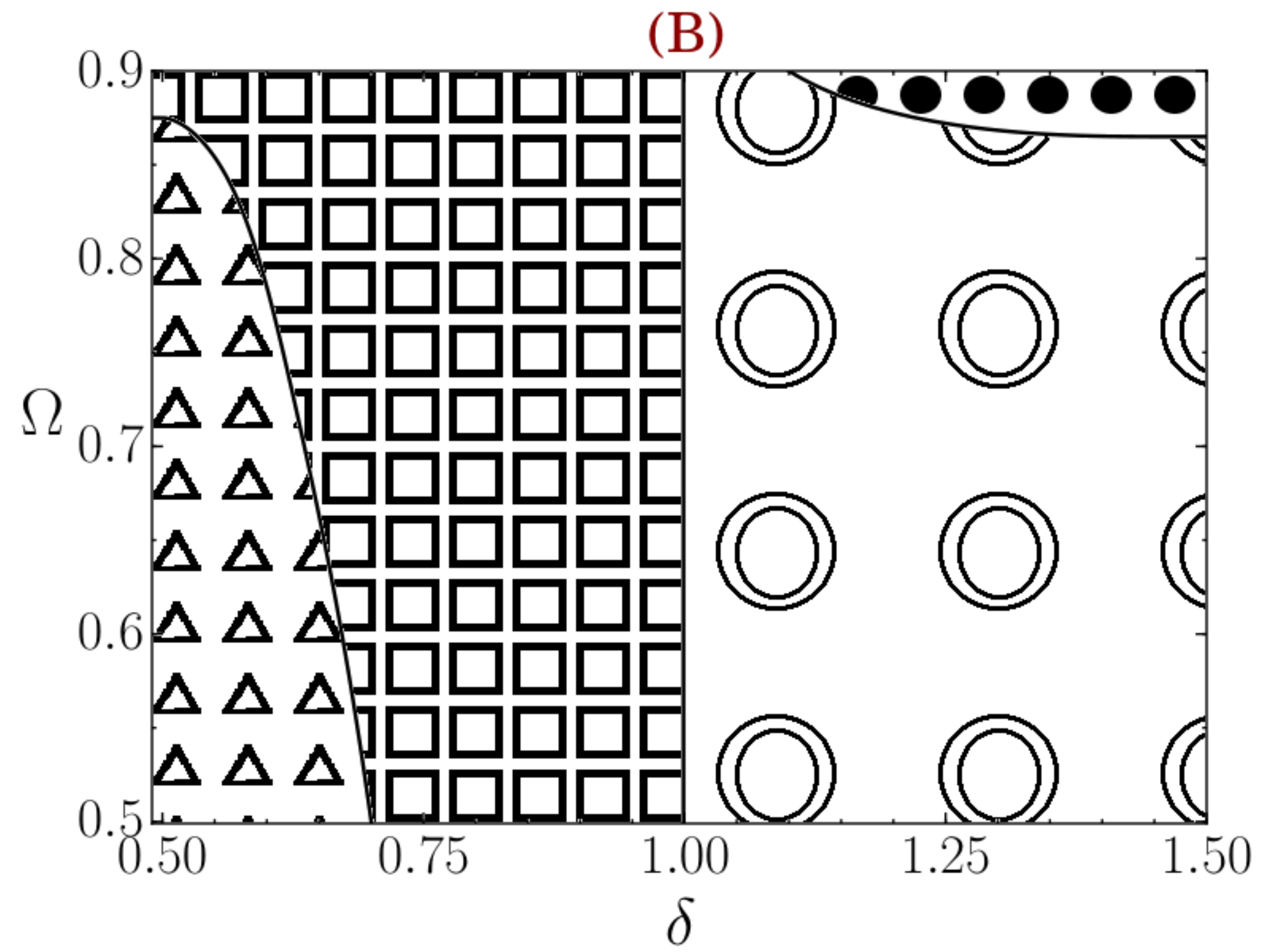}
\caption{
Diagrammatical representations of observed vortex-lattice patterns for the asymmetric $^{168}$Er-$^{164}$Dy dipolar
mixture, with the rotation frequency $\Omega$ as function of the contact interaction ratio $\delta\equiv a_{12}/a_{11}$.
Both dipoles are polarized: along the $z-$axis ($\varphi=0$) in (A); and at an angle $\varphi=60^\circ$ in (B).
The symbols filling specific intervals indicate the approximate observed vortex-lattice patterns in these regions: 
triangles for triangular-shaped, squares for squared shaped, concentric circles for circular lattices, and solid circles 
for 2D rotating droplets.}
\label{fig06}
\end{center}
\end{figure}

In this part of our study, we consider condensed system with the asymmetric-dipolar coupled species  
$^{168}$Er-$^{164}$Dy, which are confined in a pancake-shaped trap ($\lambda=50$), with the addition of a 
squared optical lattice. The parameters of the DDI in this case are given by 
$a_{11}^{(d)}=66\,a_{0} $, $a_{22}^{(d)}=131\,a_{0}$ and $a_{12}^{(d)}=a_{21}^{(d)}=94\,a_{0} $. 
The main characteristics of this system in the absence of optical lattices is that the two 
components are less miscible, as studied before in Ref.~\cite{2017jpco}.
The diagrams presented in Fig.~\ref{fig06} are summarizing the results we have obtained for the kind of vortex
patterns that are found, considering the rotation of the system, given by the parameter $\Omega$, and the 
inter- to intra-species contact interaction, $\delta$. In the diagram (A), we have considered the orientation 
of the two dipoles parallel to $z$, with $\varphi=0$, implying in a repulsive DDI. For the diagram (B), the orientation 
of the dipoles are making an angle $\varphi=60^\circ$ with respect to $z$, such that we have an attractive DDI.

As shown in the diagram (A), the vortex patterns in case of repulsive DDI are mainly with triangle shapes 
for $\delta\le 0.9$ when we assume $\Omega\lesssim 0.8$, changing to squared shapes for larger $\Omega$. 
And, for $\delta>0.9$ the patterns are mainly circular lattices, independently on the values of $\Omega$. 
Patterns with rotating droplets are observed only for larger values of $\delta$, close to $\sim 1.5$ with $\Omega$ 
close to 0.9.
In the diagram (B), we have results considering attractive DDI, with $\varphi=60^\circ$. By comparing
with the diagram (A), we noticed that the triangle patterns remain only for smaller values of $\delta$, as mostly of
the results shown with $\delta<1$ have squared pattern formats. Also, for $\delta>1$, the attractive DDI increase 
the region where we can have patterns with droplets, which can happen when $\Omega\sim 0.9$.
Sample results are given in the following three figures. 

\subsection*{\bf $^{168}$Er-$^{164}$Dy mixture, with repulsive dipole-dipole interaction}

\begin{figure}[H]
\begin{center}
\includegraphics[width=0.8\textwidth]{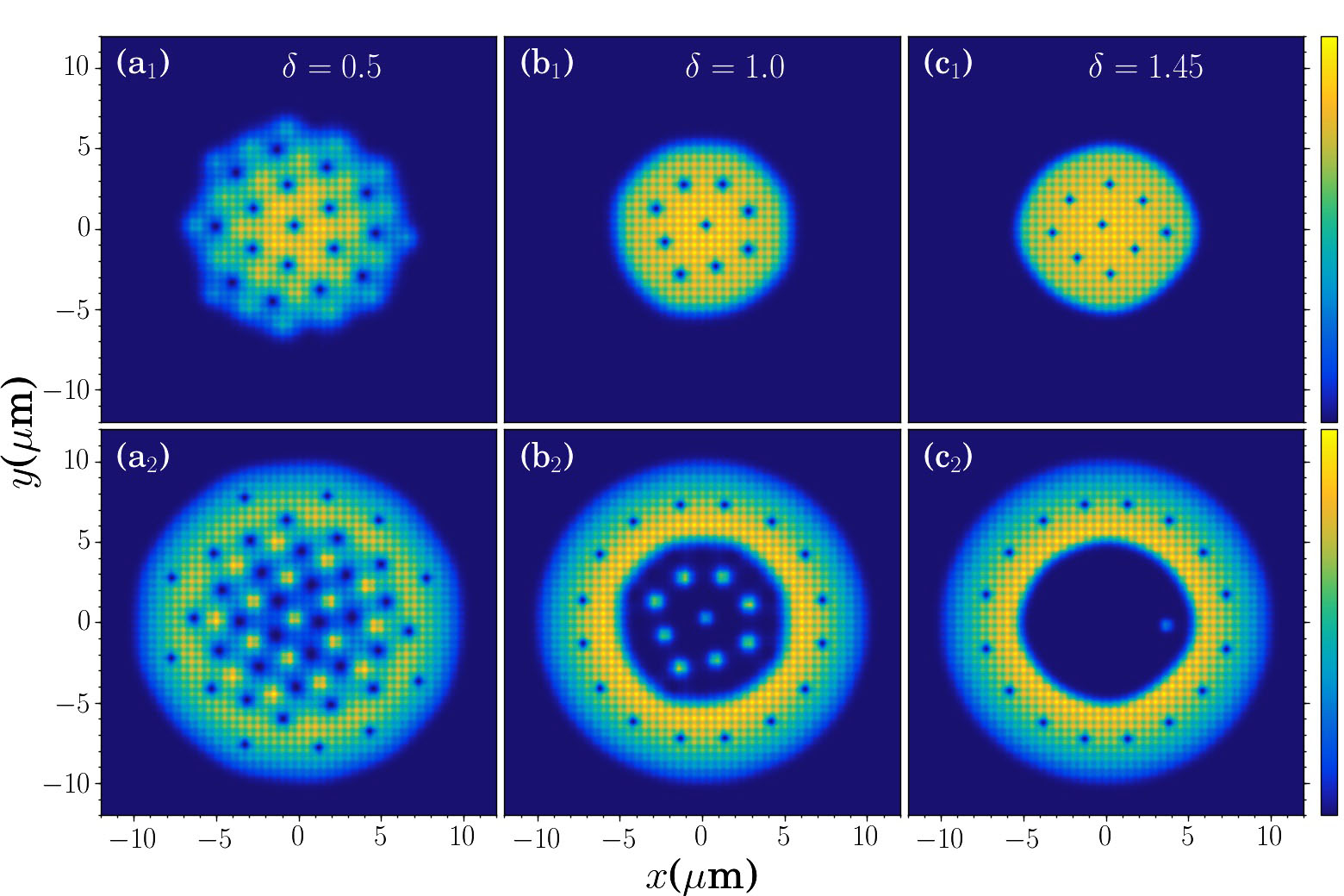}
\end{center}
\caption{2D densities of stable vortices for the dipolar mixture $^{168}$Er-$^{164}$Dy 
(when $a_{11}^{(d)}=66\,a_{0} $, $a_{22}^{(d)}=131\,a_{0}$ and $a_{12}^{(d)}=a_{21}^{(d)}=94\,a_{0} $) 
are shown in three set of panels, for $\delta$=0.5, 1.0 and 1.45 (as indicated inside the upper panels). 
The dipole orientations are parallel to $z$ ($\varphi=0$, repulsive DDI), with  the rotation parameter $\Omega =0.5$
(units $\omega_1$). The OL parameters are $V_0=15$ (units $\hbar\omega_1$) and $\pi/k=0.534$ (units $\mu$m).
Starting from zero (darker), the maximum density levels are 0.014 for (a$_1$); 0.016 for (b$_1$) and (c$_1$); and 
0.007 for the panels in the 2nd row. }
\label{fig07}
\end{figure}

The results presented in  Fig.~\ref{fig07} are for the densities of the two components, considering three values $\delta$, which
are given the ratio between inter- to intra-species scattering lengths. In this case, we are using the dipole orientations parallel
to $z-$axis ($\varphi=0$), such that we have repulsive DDI. The rotation frequency is fixed to $\Omega=0.5$.
Due to the characteristics of this coupled system, we observe that in general we have circular patterns for the densities,
where the less massive species ($^{164}$Dy, in this case) is distributed in radius larger than the other one ($^{168}$Er). 
The repulsive DDI enhances the separation between the coupled species, in particular for larger values of $\delta$ (when the inter-species
contact interaction is dominating). 
The vortex patterns are mostly triangular for $\delta<1$ and with circular ring shapes for $\delta\ge 1$.
 The $^{168}$Er-$^{164}$Dy mixture in squared optical lattice show the triangular lattice within the regime where $\delta \leq 0.9$ 
and $\Omega < 0.85$. 
 This coupled mixture, in the absence of OL, presents square vortex lattice for $\delta > 0.9$. The presence of OL 
shows different vortex phase regimes. 
The circular shaped lattices are formed due to the complete phase separation of the condensates when $\delta > 0.9$,
without being affected by the values of the rotation parameter $\Omega$.

\begin{figure}[H]
\begin{center}
\includegraphics[width=0.6\textwidth]{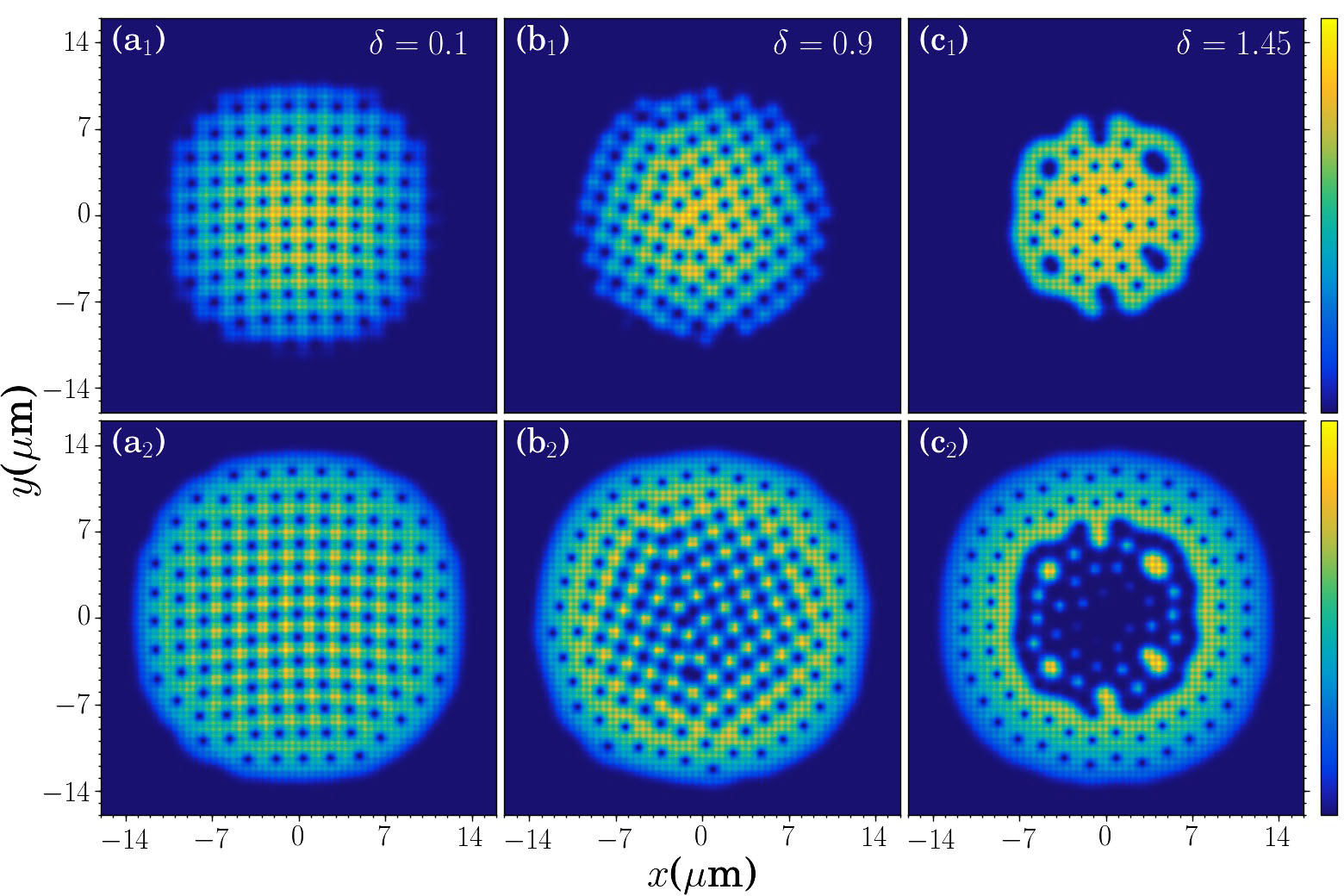}
\end{center}
\caption{
Effect of increasing the rotation frequency with $\Omega=0.9$ in the density patterns shown in 
Fig.~\ref{fig07} for the $^{168}$Er-$^{164}$Dy mixture, by considering the 
contact-ratio $\delta=$0.1, 0.9 and 1.45 (as indicated inside the upper panels).
Other parameters are the same as in  Fig.~\ref{fig07}.
Starting from zero (darker), the maximum density levels are 0.0073 for (a$_1$) and (b$_1$); 
0.0090 for (c$_1$); 0.0045 for (a$_2$) and (b$_2$); and 0.0058 for (c$_2$).
} 
\label{fig08}
\end{figure}

With the results presented in Fig.~\ref{fig08}, we verify how the results given in Fig.~\ref{fig07} are affected by increasing the 
rotation. Therefore, we increase to 0.9 the value of $\Omega$ (in Fig.~\ref{fig07} is 0.5).
 As one can noticed, by increasing the rotation the coupled system becomes more miscible, such that
for $\delta>1$ the circular patterns are not well defined as in Fig.~\ref{fig07} (see, for example, Fig.11 of
Ref.~\cite{Kumar2017}). 
We can also verify the increasing number of rotating droplets and density peaks, which are formed due to multiple vorticity, as also 
discussed in the case of Fig.~\ref{fig03} for the $^{164}$Dy-$^{162}$Dy mixture.

\subsection*{\bf $^{168}$Er-$^{164}$Dy mixture, with attractive dipole-dipole interaction}

\begin{figure}[H]
\begin{center}
\includegraphics[width=0.6\textwidth]{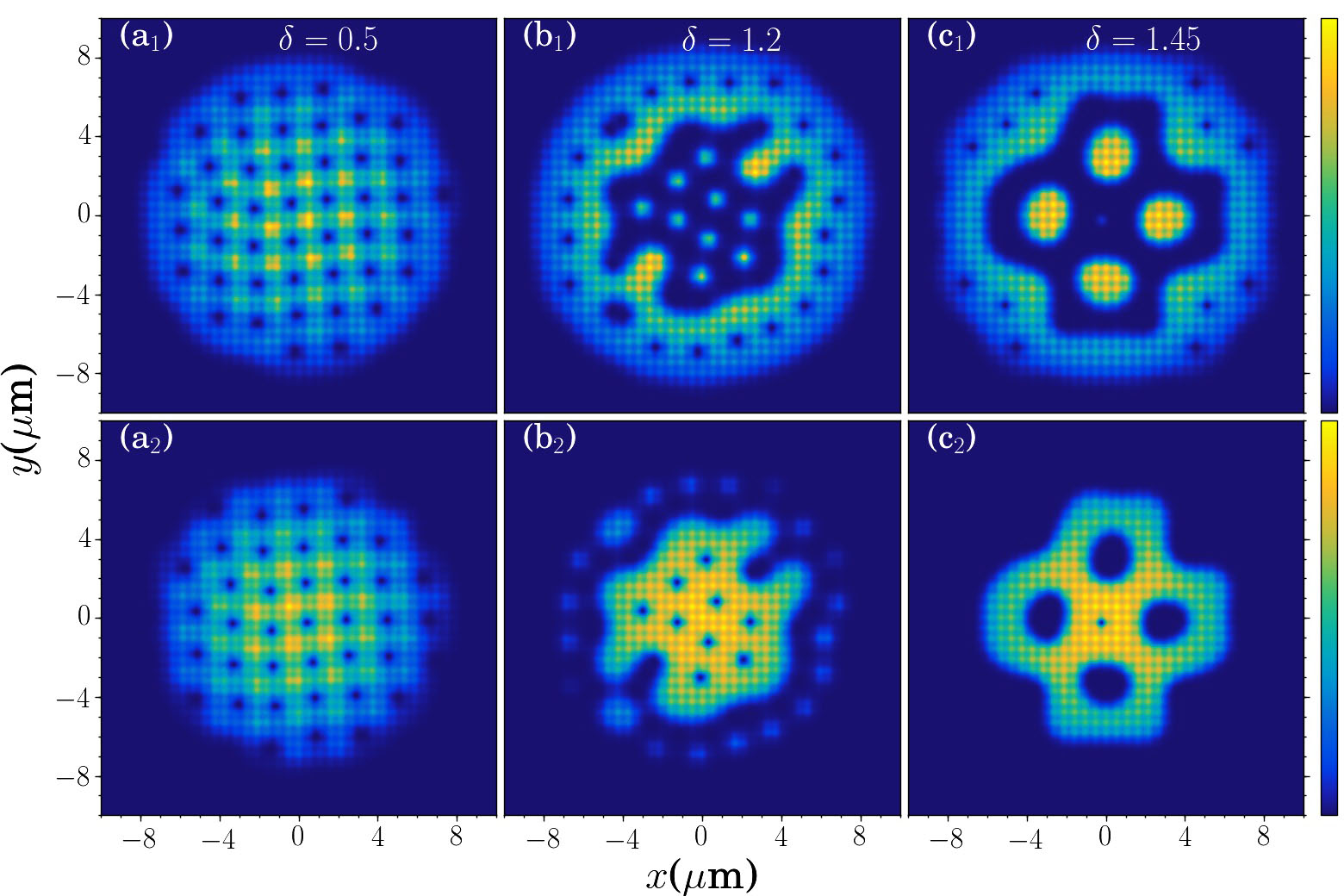}
\end{center}
\vspace{-0.5cm}
\caption{
 2D density patterns are shown for the coupled mixture $^{168}$Er-$^{164}$Dy 
 in three set of panels, with $\delta$ from 0.5, 1.2 and 1.45 (as indicated inside the upper frames), 
considering $\Omega=0.9$ (as in  Fig.~\ref{fig08}) and the dipole orientations tuned to $\varphi=60^\circ$ 
(attractive DDI). Other parameters are as in Figs.~\ref{fig07} and \ref{fig08}. 
The density levels are from zero to 0.019 for (a$_1$); 0.018 for (b$_1$); and 0.020 for the other panels.
}\label{fig09}
\end{figure}

Next, in Fig.~\ref{fig09}, we change the orientation of the dipoles from $\varphi=0$ (used in Fig.~\ref{fig08}) to 
$\varphi=60^\circ$, such that the DDI becomes attractive. The effect of more attraction, and also by keeping the rotation 
high with $\Omega=0.9$, is that the radius is reduced, as well as the number of vortices, with 
different vortex patterns being observed.

\subsection{\bf $^{164}$Dy-$^{52}$Cr and $^{164}$Dy-$^{87}$Rb dipolar mixtures in squared optical lattice}

\begin{figure}[H]
\begin{center}
\includegraphics[width=0.7\textwidth]{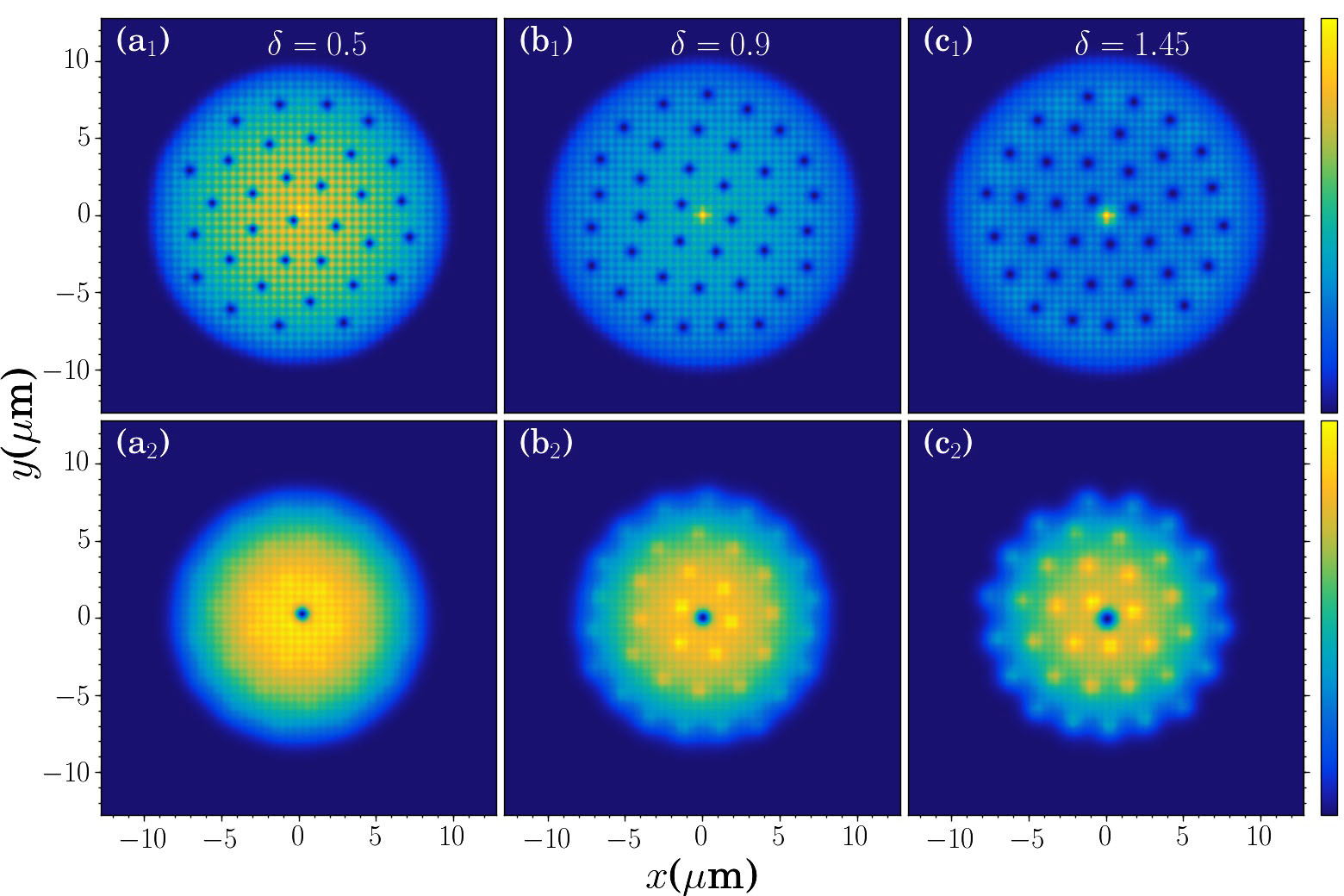}
\end{center}
\vspace{-0.5cm}
\caption{
2D densities $|\protect\psi_{j=1,2}|^{2}$ of stable vortices for the dipolar mixture $^{164}$Dy-$^{52}$Cr  
(when $a_{11}^{(d)}=131\,a_{0} $, $a_{22}^{(d)}=16\,a_{0}$ and $a_{12}^{(d)}=a_{21}^{(d)}=25\,a_{0} $) 
are shown for $\delta$=0.5, 0.9 and 1.45 (as indicated inside the upper frames), with $\Omega=0.5$
and for repulsive DDI with $\varphi=0^\circ$.
The OL and other parameters are as in Fig.~\ref{fig07}.  
 The density levels are from zero to 0.010 for (a$_1$); 0.011 for (b$_1$); 0.013 for (c$_1$); and 
 0.009 for the other panels.
}
\label{fig10}
\end{figure}
\begin{figure}[H]
\begin{center}
\includegraphics[width=0.7\textwidth]{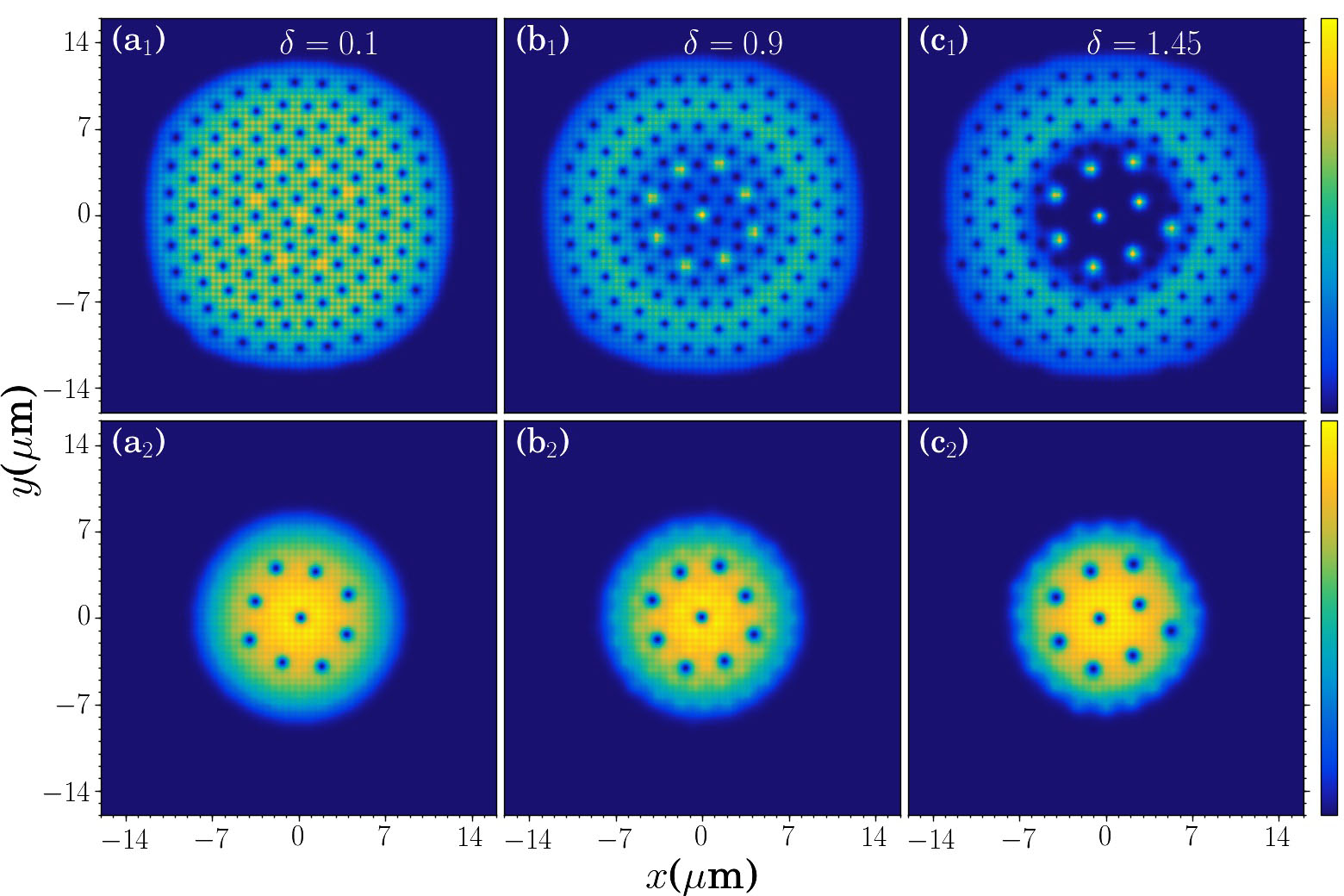}
\end{center}
\caption{
Effect of changing the rotation frequency to  $\Omega=$0.9 in the density patterns as given in
Fig.~\ref{fig10}, for the $^{164}$Dy-$^{52}$Cr dipolar mixture with $\varphi=0^\circ$ 
(Other parameters are as in Fig.~\ref{fig10}).  
 Starting with zero (darker), the maximum density levels are 0.0130 for (a$_1$); 0.0064 for (b$_1$); 
 0.0078 for (c$_1$); 0.0080 for (a$_2$) and (b$_2$); and 0.0090 for (c$_2$).
}
\label{fig11}
\end{figure}

\begin{figure}[H]
\begin{center}
\includegraphics[width=0.7\textwidth]{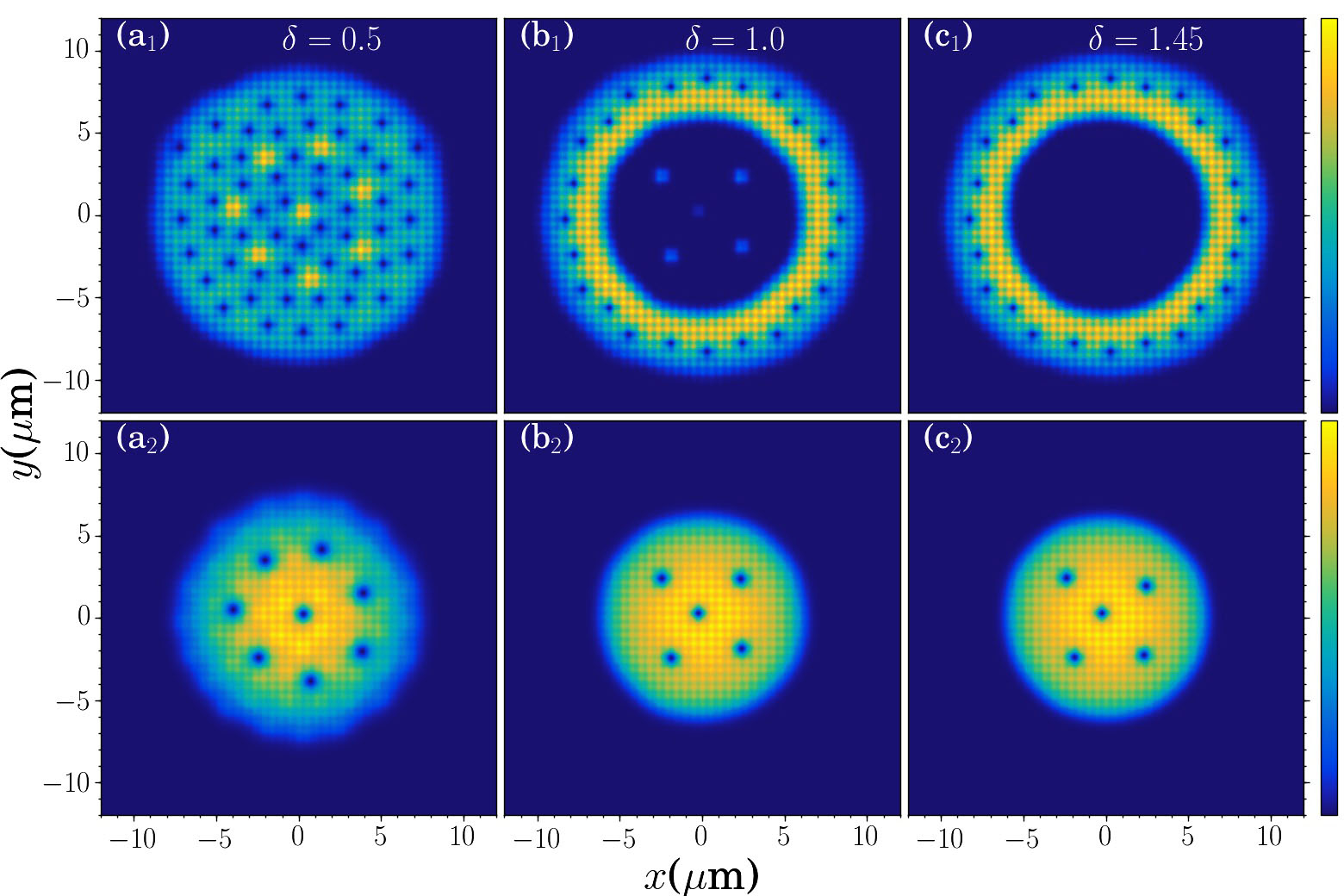}
\end{center}
\caption{
The 2D component densities of the mixture $^{164}$Dy-$^{52}$Cr are shown 
 with the same parameters as in Fig.~\ref{fig10}, but with $\Omega=$0.9  
 and $\varphi=60^\circ$ (attractive DDI).
Starting from zero (darker), the maximum density levels are 0.011 for the 1st
row and 0.012 for the 2nd row of panels.
}
\label{fig12}
\end{figure}

\begin{figure}[H]
\begin{center}
\includegraphics[width=0.6\textwidth]{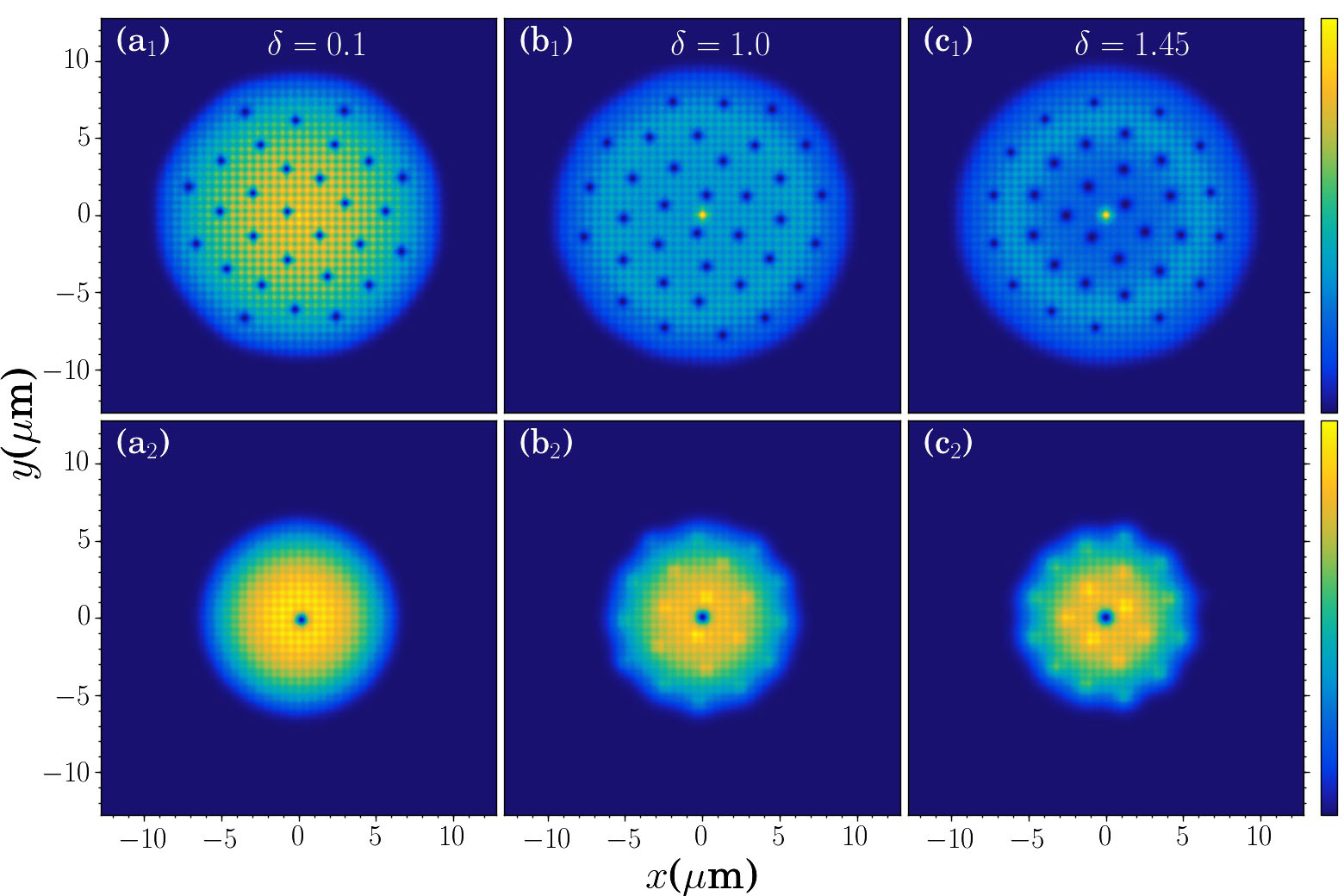}
\end{center}
\caption{
The 2D densities of stable vortices for the dipolar mixture 
$^{164}$Dy-$^{87}$Rb (when $a_{11}^{(d)}=131\,a_{0} $, with 
$a_{22}^{(d)}$ and $a_{12}^{(d)}$ negligible) are shown for $\delta$=0.1, 1.0 and 1.45 
(as indicated inside the upper frames), with $\Omega=0.5$, and considering repulsive DDI
($\varphi=0^\circ$). The other parameters are as in Fig.~\ref{fig11}.  
Starting from zero (darker), the maximum density levels are 0.010 for the 1st row
and 0.009 for the 2nd row of panels.}
\label{fig13}
\end{figure}

In this subsection we consider the coupled mixtures which have more stronger asymmetry,  given by 
$^{164}$Dy-$^{52}$Cr and $^{164}$Dy-$^{87}$Rb. Besides their dipole properties, these systems have also 
larger mass difference than the ones considered before. 
In these cases, considering the dysprosium as the first component, the corresponding parameters of the DDI are  
$a_{11}^{(d)}=131\,a_{0} $, $a_{22}^{(d)}=16\,a_{0}$ and $a_{12}^{(d)}=a_{21}^{(d)}=25\,a_{0} $
for $^{164}$Dy-$^{52}$Cr; with $a_{22}^{(d)}$ and $a_{12}^{(d)}$ almost zero for $^{164}$Dy-$^{87}$Rb. 
We choose to give more details on the mixture with chromium, as in this case we have large asymmetry 
but with all non-zero dipolar parameters. The vortex-pattern results are shown through Figs.~\ref{fig10}, 
\ref{fig11} and \ref{fig12}, in analogy with the previous cases. With Fig.~\ref{fig13}, we conclude this 
section with the corresponding vortex-pattern results for the densities obtained 
for $^{164}$Dy-$^{87}$Rb mixture, in which just one of the intra-species dipolar parameters is non-zero.

In each of the figures we display a set of three panels, considering three characteristics values of the parameter 
$\delta$, for fixed rotations and dipole orientations. 
We start in Fig.~\ref{fig10}, by considering $\Omega=0.5$ and $\varphi=0$ (repulsive DDI). As verified, 
the density and number of vortices are distributed in a larger radius higher than 10 for the first component,
the dysprosium, with the chromium density distributed in a smaller radius. In this case, this distribution of
the density (in a larger space for the more massive species) is related to the repulsive intra-species dipolar 
interaction for the first component, $a_{11}^{(d)}=131\,a_{0} $, which is about 8 times larger than the 
intra-species interaction of the second component, $a_{22}^{(d)}=16\,a_{0} $;  and about 5 times larger 
than the inter-species interaction, $a_{12}^{(d)}=25\,a_{0} $.  As also shown in this figure, the number of vortices in 
the second component is more affected by the changes in the inter- to intra-species parameter, increasing for $\delta>1$.

Next, in Fig.~\ref{fig11}, we increase the rotation parameter to $\Omega=0.9$, keeping the same orientation 
of the dipoles parallel to $z$ (repulsive DDI). As expected the radial distribution of the densities increases,
in particular for the first component, where we also notice the increasing in the number of vortices. 
The maximum localization of the distribution of this first component moves outside the center as the 
inter-species contact interaction is increased.
For the second component, which has the density distributed more close to the center, we notice that
the effect of an increasing rotation has the effect to increase the number of vortices, which is clear for smaller values 
of $\delta$. However, by increasing $\delta$, with such $\Omega=0.9$, the vortices
merge together, as seen in the three panels.

With Fig.~\ref{fig12}, we show the effect in the vortex-patterns obtained for the $^{164}$Dy-$^{52}$Cr dipolar mixture,
by changing the DDI to an attractive one, with $\varphi=60^\circ$.
In this case, we keep the same rotation parameter as in Fig.~\ref{fig11}. By changing to attractive the DDI, 
the two densities become more concentrated, with almost all the density corresponding to the first component 
distributed in a ring around the 2nd component. 

Finally, we present sample results of the densities for stable vortex-lattice distributions considering the 
$^{164}$Dy-$^{87}$Rb dipolar mixture, in three panels of the Fig.~\ref{fig13}. As this case is similar to the case 
that we have chromium instead of the rubidium, with larger asymmetry in the dipolar properties of the two 
species, we consider a case where the other parameters (except the dipolar ones) are the same as the 
ones considered in Fig.~\ref{fig10}, with repulsive DDI ($\varphi=0$) and rotation given by $\Omega=0.5$.
As in the other cases, $\protect\delta$ is varying from 0.5 to 1.45. 

The net difference between the two cases in this comparison is the decreasing in the vorticity.
The comparison between the results obtained for coupled dipolar mixtures with one of the species having 
a strong dipole strength, the $^{164}$Dy, with two cases where the second component have weak or zero 
dipole strengths, such as $^{52}$Cr and $^{87}$Rb, is quite indicative of the behavior of mixtures when
we have a non-dipolar component.

\section{Summary and Conclusions}

In our study of structural vortex-lattice transitions with binary mixtures of dipolar BECs trapped in strong 
pancake-type symmetry and loaded in rotating squared optical lattice, we consider the following mixtures 
of dipolar isotopes: $^{164}$Dy-$^{162}$Dy, $^{168}$Er-$^{164}$Dy, $^{164}$Dy-$^{52}$Cr and
$^{164}$Dy-$^{87}$Rb.
As the strength of the corresponding dipolar interactions are fixed for these systems, we consider two possible 
orientations of both aligned dipoles, by tuning the polarization angle to $\varphi=0$ or $60^\circ$ with respect to 
the $z-$axis. By using these two orientations, we consider repulsive dipole-dipole interactions ($\varphi=0$), 
as well as attractive ones  ($\varphi=60^\circ$). Once considered the intrinsic dipolar properties of the binary 
mixture, together with the trap conditions (pancake-type with aspect ratio $\lambda=50$) and optical lattice 
parameters, the other main physical variables in our study to produce 
different lattice-vortex patterns are  the ratio between inter- to intra-species two-body interaction $\delta$ 
and the rotation frequency $\Omega$ (in units of the trap frequency $\omega_1$).

By considering the kind of patterns which are obtained in our numerical investigation, for different pair of
parameters $\Omega$ and $\delta$, all the results for the symmetric $^{164}$Dy-$^{162}$Dy and asymmetric 
$^{168}$Er-$^{164}$Dy dipolar mixtures are summarized in two diagrams given in Figs.~\ref{fig01} and \ref{fig06}, 
respectively. The diagrams are followed by specific examples of vortex-pattern results. 
For the other binary mixtures, we just choose representative results of the vortex-lattice structures.
In the specific results, which are being displayed along this work, we consider 
two representative values for the rotation parameter, $\Omega=$0.5 and 0.9, corresponding to low- and high-rotation speed. 
For the squared optical lattice, the strength is fixed to $V_0=15$ with the lattice grid given by $\pi/k=0.534$ (units of
$\mu$m), suggested by existing experimental possibilities. The lattice parameter $k$ was varied in the specific
case of symmetric-dipolar mixture, showing a way to control the vortex-lattice patterns 
and indicating the grid interval where the OL potential is dominant.  
Besides that, the vortex-lattice structures are changed by varying the inter-species contact interactions,
with the assumption that the corresponding intra-species interactions are kept equal for both species.

The symmetric $^{164}$Dy-$^{162}$Dy dipolar mixture was first analyzed, as it is more appropriate
to verify the optical lattice effect, in both the cases that we have repulsive or attractive DDI, by tuning the dipole 
orientations $\varphi$.
As compared with studies having no OL potential, we noticed that the addition of a  
squared OL potential (in the plane defined by the pancake-type trap) has the effect to change triangle
vortex-lattice patterns  to squared formats when $\delta \lesssim 1$. 
However, for $\delta\ge 1$, it is observed a structural transition in the patterns, with the OL  
reinforcing patterns with stripes, with domain walls and vortex sheets. 
In particular, interesting 2D rotating droplet vortex structures are observed by increasing the asymmetry between 
inter- and intra-species interaction (such as we have exemplified, with $\delta\sim 1.45$).
A relative shallow OL facilitates the appearance of droplet vortex structures in some particular regimes 
of the parameters $\delta$ and $\Omega$. For the strength of the OL we select $V_0=15$, after verifying
that the vortex-lattice structures are in general similar. 
We have also observed that droplets lattice patterns are not supported with deeper OL such as with $V_0 > 25$,
when considering rotation frequency not high, near $\Omega=0.5$. As increasing the frequency, with 
$\Omega > 0.65$, more vortices are established, which start to connect together forming domain-wall structures, 
with droplet vortex patterns no more being verified.
 
We change the DDI strength from repulsive (when $\varphi=0$) to attractive one, by tuning the dipole orientation 
to $\varphi=60^\circ$. In this case, anisotropic effects due to the dipolar interaction are visualized with the radius 
of the condensed mixture being significantly reduced in relation to the case that $\varphi=0$ (see Figs.~\ref{fig04} 
and \ref{fig05}). When the DDI becomes attractive, we also note that the effect of OL is less pronounced. 
The attractive two-body interactions, either by contact or DDI, reduce the number of vortices, with 
the phase diagram becoming similar to the case of dipolar mixtures in harmonic traps. 
The main different characteristics is that the OLs support rotating droplet vortex structures for 
$\delta>1.3$ and $\Omega>0.875$. As the interactions are attractive, the number of vortices is not 
enough to create droplets for $\Omega<0.875$. 
In the regime with $\delta>1.3$ and $\Omega<0.875$, we have only separated phase mixtures  
with few vortices. 
 
Next, we study the asymmetric-dipolar mixture $^{168}$Er-$^{164}$Dy, which is verified to be less 
miscible in previous studies where there is no OL interactions. 
Again we consider two phase diagrams in a plane defined by $\Omega$ and $\delta$, given in Fig.~\ref{fig06},
for summing up our results for the lattice-vortex patterns. 
As verified in such a case, there is no 
relevant effect verified in the vortex-lattice patterns by varying the rotation frequency $\Omega$, which 
are mainly squared vortex lattices for $\delta<1$, with circular vortex lattice for $\delta>1$. 
On the other hand, after applying the squared OL on this asymmetric mixture, we can verify that the
vortex-lattice patterns are more affected by changing the rotation frequency. 
As indicated in the $\delta-\Omega$ phase diagram presented in the left frame of Fig.~\ref{fig06}, 
for the case that the polarized dipoles are aligned with the $z-$axis ($\varphi=0$), when 
$\delta<0.9$, the vortex-lattice patterns produced are mainly with triangular formats  when $\Omega<0.85$, 
changing to squared formats when $\Omega>0.85$. For $\delta>0.9$ the vortex-lattice structures are
predominantly with concentric circles, with 
rotating droplet vortex-lattice patterns being verified in the limited regime for large values of $\delta$ ($\sim$1.45)
and $\Omega$ ($\sim$0.9). Sample illustrative vortex-lattice patterns are shown in Fig.~\ref{fig09}.
Further, by changing the DDI to be attractive, with $\varphi=60^\circ$, as shown in the right frame of the
Fig.~\ref{fig06}, for $0.7<\delta<1$, most of the patterns change from triangular (when the DDI are repulsive) 
to squared formats (when the DDI are repulsive), except for smaller values of $\delta$ where the triangular
formats remain.
Droplet vortex structures can also be observed in the high rotation regime, where $\Omega$ is close to 
0.9, when $\delta>1.2$. See, for example, the patterns (c$_i$) shown in Fig.~\ref{fig08}, in this case.

More briefly we consider the highly asymmetric mixtures, with larger differences in the mass and dipole 
moments, as $^{164}$Dy-$^{52}$Cr and $^{164}$Dy-$^{87}$Rb. These mixtures show only triangular and 
half-quantum vortex lattices. The dipole moment of $^{164}$Dy is comparatively much larger than the dipole 
moments of the other component. So, the first component ($^{164}$Dy) produces more vortices than the second one. 
Besides that, we have observed that both $^{164}$Dy-$^{52}$Cr and $^{164}$Dy-$^{87}$Rb present similar 
characteristics.   

In our present investigation, we have studied dipolar-symmetric and dipolar-asymmetric binary systems, which are 
confined in stable pancake-shaped configurations with fixed optical lattice parameters, where most of the results are shown
for two moderate low and high rotation frequencies ($\Omega=$0.5 and 0.9). Apart of that, for the vortex-pattern structures, 
the inter- to intra-species repulsive contact interactions $\delta$ and the polarization angle of the DDI $\varphi$ are the main 
relevant parameters to be considered.
By increasing $\delta$ the miscibility of the two components is reduced, implying in vortex-pattern structures having different 
domain-wall shapes in the dipolar-symmetric case; and in radial space separations of the components in the 
dipolar-asymmetric case.
The polarization angle can change the system from repulsive to attractive, by going to larger angles, which will reduce the 
radius, having relevant effects on the pattern structures, particularly for larger $\delta$, in the interplay between repulsive
contact interactions (increasing the immiscibility) and the attractive DDI (reducing the radial distribution of the densities).

The optical lattice parameters are adjusted when considering possible experimental setups,
as well as to not affect strongly the behavior of other relevant quantities we are studying.
A specific example is presented in Fig.~\ref{fig03}, where we show how
the OL can affect the vortex-lattice patterns by changing the OL grid-spacing parameter.
As observed, for values of $k$ near zero (very large grid spacing), there is almost no oscillations along the trap, with practically no 
OL effect. Also, in the other limit, for large values of $k$ (very small grid spacing), the trap is practically not affected, because the 
oscillations will average to a constant. However, for intermediate values of $k$, we can have dramatic changes of patterns with further 
loss of cylindrical symmetry due to alignment of the lattice and corresponding BEC localization. 

Finally, we understand that the present results can be useful to calibrate going on experiments with dipolar mixtures, 
as the recent ones, $^{164}$Dy-$^{162}$Dy and $^{168}$Er-$^{164}$Dy, which are under active investigations,
when considering pancake-type trap symmetries and loaded in squared optical lattices. 

\ack 
We thank the Brazilian agencies 
FAPESP - Funda\c c\~ao de Amparo \`a Pesquisa do Estado de S\~ao Paulo (Procs. 2014/01668-8, 2016/17612-7 
and 2017/05660-0), 
CNPq - Conselho Nacional de Desenvolvimento Cient\'\i fico e Tecnol\'ogico (Procs. 306191/2014-8 and 
304468/2014-2) and
CAPES - Coordena\c c\~ao de Aperfei\c coamento de Pessoal de N\'\i vel Superior (LT), for partial financial support.

\end{document}